# Handedness of magnetic-dipolar modes in ferrite disks

E.O. Kamenetskii

Department of Electrical and Computer Engineering,
Ben Gurion University of the Negev, Beer Sheva, 84105, ISRAEL

**Absract**

For magnetic-dipolar modes in a ferrite, components of the magnetic flux density in a helical coordinate system are dependent on both an orientation of a gyration vector and a sign of a pitch. It gives four types of helical harmonics for magnetostatic-potential wave functions in a ferrite disk. Because of the reflection symmetry breaking, coupling between certain types of helical harmonics takes place in the reflection points. The reflection feature leads to exhibition of two types of resonances: the "right" and "left" resonances. These resonances become coupled for a ferrite disk placed in a homogeneous tangential RF magnetic field. One also observes such resonance coupling for a ferrite disk with a symmetrically oriented linear surface electrode, when this ferrite particle is placed in a homogeneous tangential RF electric field. In a cylindrical coordinate system handedness of magnetic-dipolar modes in a ferrite disk is described by spinor wave functions.

PACS: 76.50.+g; 68.65.-k; 11.30.Er

# 1. INTRODUCTION

In a ferromagnet, there is a long-ranged dipole-dipole interaction between the magnetic moments of the atoms. This interaction is considered as purely relativistic in origin and is supposed to be as an additional factor to the short-ranged exchange interaction, which is the strongest interaction between atoms in a ferromagnet. In theoretical studies of bulk magnetic materials, the dipolar interaction is often ignored [1].

In quasi-two-dimensional systems, the dipolar interaction can play an essential role in determine the magnetic properties. In these systems the short-range exchange interactions alone are not necessarily sufficient to establish a ferromagnetically ordered ground state. The dipolar interaction is important in stabilizing long-range magnetic order in two-dimensional systems, as well in determining the nature and morphology of the ordered states. Another important property of the dipolar interaction for two-dimensional films is that it breaks the symmetry between the out-of-plane orientation of the spins and the in-plane orientation of the spins in the ordered state [2]. The long-range nature and the anisotropic character of the dipolar interaction require considerable care in the evolution of the dipolar contribution to the magnetic energy. A generalized form of the Ewald summation technique provides a powerful means of dealing with the slowly convergent nature of dipolar sums. The basis of the summation technique is the separation of the interaction into a localized part and a long-range part [2-4].

In the analysis of the spin-wave spectra in two-dimensional spin systems, including the dipolar interaction considerably complicates the problem. There has been developed the microscopic formalism for the dipole interactions in spin-wave ferromagnetic films [3]. On the other hand, there has been extensive work to generalize the magnetostatic-wave (or continuous-medium) theory to include the exchange effects in two-dimensional ferrites [5]. These complicated dipole-exchange theories should be applicable, however, for ultrathin ferrite films. In microwave experiments on the exchange effects, there are the films with thickness not more than units of micrometers. For films with thickness about tens of micrometers one successfully uses the continuum approach and describes the magnetization dynamics based on the susceptibility and permeability tensors. In this case the magnetic stiffness is characterized by macroscopic magnetization and the problem is solved based on the magnetostatic-wave (MS-wave) theory [6, 7].

The fields associated with the various degrees of freedom in a crystal have been quantized. For example, magnons are the elementary excitation of an exchange-coupled spin system. The canonical field variables associated with each normal spin-wave mode are determined and expressed in terms of creation and annihilation operators using the Holstein-Primakoff transformation [8]. The modes are taken to be plane waves. In unbounded ferromagnetic films one has properties of translational symmetry for a crystal lattice. In this case the summation technique taking into account the long-range character of the dipolar interaction is applicable [3]. The procedure can be extended for two-dimensional artificial lattices with magnetic particles interacting through the long range dipolar forces [9]. It can also be supposed that certain conditions of translational symmetry in a plane of a film might be introduced for an infinite ferrite strip. Situation, however, becomes strongly different in a case of a thin-film *ferrite disk*. Because of lack of in-plain translational symmetry for the long-range dipolar interaction in a disk-film ferrite sample, it becomes difficult to use the summation technique and perform the Holstein-Primakoff transformation. (It is necessary to point out here that nonlinear 2D spin-dynamics processes in large systems with circular symmetry, such as the vortex-magnon coupling [10], are beyond the scope of the present consideration.)

Diagonalization of energy of linear magnetic-dipolar continuous-medium modes with the motion-equation description is possible, however, for a disk-film ferrite samples with a small thickness-to-diameter ratio [11]. For such samples the oscillating energy becomes diagonalized in consideration of the *reflexively-translational* motion of "flat" quasiparticles – the light magnons [12]. One has situation very similar to the dipole-interaction "flat" quasiparticles (electron-hole pairs – excitons) in disk-form semiconductor dots [13]. It is necessary to note that energy diagonalization for magnetic-dipolar modes is possible only for a normally magnetized ferrite disk. In this case for spin $\vec{S}$ precessing in a DC magnetic field $\vec{H}_0$ the energy has a form $\vec{S}\cdot\vec{H}_0$ while the equation of motion is $\dot{\vec{S}} \propto \vec{S}\times\vec{H}_0$. So for oppositely precessing spins in magnetic-dipolar "flat" modes the energy eigenvalues are all positive [14].

Multi-resonance magnetic-dipolar [or magnetostatic (MS)] oscillations in ferrite disks were experimentally investigated, for the first time, by Dillon [15] and then analyzed more in details by Yukawa and Abe [16]. In a view of recent experiments of a microwave magnetoelectric (ME) effect [17], an interest in MS oscillating spectra of a normally magnetized ferrite disks was renewed. The *macroscopically quantum analysis* [11, 12], being a new consideration of an old topic, touches upon fundamental aspects of the physics of MS oscillations and, hopefully, gives a clue to the microwave ME effect in ferrite particles. In microwaves, ferrite resonators with multi-resonance MS-wave oscillations may have characteristic sizes two-four orders less than the free-space EM wavelength at the same frequency and, at the same time, much more than the exchange-interaction wavelength. So MS-wave oscillations in a ferrite sample occupy an intermediate position between the "pure" electromagnetic and spin-wave (exchange) processes. The energy density of the MS-wave oscillations is not the electromagnetic-wave density of energy and not the exchange energy density as well. These "microscopic" oscillating objects – the particles – may interact with the external EM fields by a very specific way, forbidden for the classical description.

In the existing MS-wave theory [6, 7], MS-potential function $\psi$ is described by the Walker equation:

$$\nabla \cdot (\vec{\mu} \cdot \nabla \psi) = 0 \qquad (1)$$

inside a ferrite (here $\vec{\mu}$ is the permeability tensor) and by the Laplace equation:

$$\nabla^2 \psi = 0 \qquad (2)$$

outside a ferrite region. Both these equations are separated in a cylindrical coordinate system. In a normally magnetized ferrite disk with a small thickness-to-diameter ratio one can successfully use separation of variables for the MS-wave function in a cylindrical coordinate system [11]. In this case the spectrum is found as a result of a solution of a system of two characteristic equations for MS waves in a normally magnetized ferrite slab [18] and in an axially magnetized ferrite rod [19]. The MS-potential wave functions are represented by sets of the "thickness" and "in-plane" (or "flat") modes and are described by the Schrödinger-like equation. In this case, one obtains the normalized spectrum of energy eigenstates [11, 12].

An analysis of energy spectra of MS oscillations shows, however, that modes with different signs of an azimuth number may have different energy levels [20]. The fact that the solution of the problem is dependent on a sign of an azimuth number reveals a contradiction in formulation of the energy orthonormality relations. It means that MS-wave functions cannot be considered as single-valued functions. At the same time, following axioms of non-relativistic quantum mechanics [21], each state function, as well as a superposition of the state functions *must be a single-valued analytic expression* satisfying the boundary conditions for the given system. In our problem the ambiguity arises from boundary conditions on a lateral surface of a ferrite disk. An analysis shows that MS oscillations being characterized by a discrete spectrum of energy levels have also specific surface magnetic currents, which are described by double-valued functions and cause the parity-violating perturbations. Because of such magnetic currents, there are the parity-odd, time-reversal-even motion processes having a clear analogy with the anapole-moment characteristics in the weak interaction [22].

To understand the problem why multiple valuedness in spectra of MS oscillations in a ferrite disk takes place, we suggest here to use the helical coordinate system together with the cylindrical one. In a helical coordinate system the azimuth number does not necessarily have to be an integer as it would be in a standard cylindrical system. Unlike the Cartesian or cylindrical coordinate systems, in the helical system, two different types of solutions are admitted, one right-handed and one left-handed. In the helical coordinate system solutions of Eqns. (1) and (2) are not separable. Since the helical coordinates are nonorthogonal and curvilinear, different types of helical coordinate systems can be suggested. In our analysis we will use Waldron's coordinate system [23]. As an alternative coordinate system, we can point out, for example, the system proposed by Lin-Chung and Rajagopal [24]. Waldron showed [23] that the solution of the Helmholtz equation in a helical coordinate system can be reduced to the solution of the Bessel equation. With use of the Waldron coordinate system, Overfelt had got analytic exact solutions of the Laplace equation in a helical coordinate system with a reference to the helical Bessel functions and helical harmonics for static fields [25].

In this paper we analyze helical harmonics for MS-wave oscillations in thin-film ferrite disks. Because of the helical coordinates, matching across the boundaries (the boundary plates $z = 0, d$ and the lateral-surface, $r = \Re$, boundary) leads to a nontrivial but solvable problem. Two factors play an essential role in our case. There is a gyrotropic (off-diagonal) term in the permeability tensor and the presence of two turn points on boundary plates ($z = 0, d$) in a ferrite disk. Since helical coordinates are not separable, to get the physically adequate models for MS oscillations in a ferrite disk we have to correlate the obtained results with the ones given from the cylindrical coordinate system. If in a structure under consideration one does not distinguish the left- or right-handedness, the results obtained in helical coordinates will be the same as in cylindrical coordinates. When a structure demonstrates the handedness properties, the physical models in cylindrical and helical coordinates will be different. As it will be shown, in a ferrite disk resonator there exist *four helical harmonics* for the MS-potential function $\psi$. So the wave function $\psi$ must have *four components*, which can be combined to form a single-column matrix:

$$\psi(\vec{r},t) = \begin{pmatrix} \psi_1(\vec{r},t) \\ \psi_2(\vec{r},t) \\ \psi_3(\vec{r},t) \\ \psi_4(\vec{r},t) \end{pmatrix}. \tag{3}$$

This implies existence of four simultaneous first-order partial differential equations that are linear and homogeneous in the four $\psi$'s.

Our analysis results in the Dirac-like quasiparticle spectrum. At present, in different 2D condensed matter systems "relativistic" Dirac-like spectrum of quasiparticle excitations becomes a subject of a special attention (see e.g. [26] – [28]). Understanding the nature of such quasiparticle energy states is of considerable importance. In that sense, magnetic-dipolar-mode oscillations are ideal since they have very long wavelength and are easily investigated by experimental techniques.

The paper is organized as follows. In Sections 2 and 3 we analyze the main properties of magnetic-dipolar modes, such as the spectra, excitation problems and Faraday rotation, and show that, physically, these modes occupy a specific place between the "pure" electromagnetic and spin-wave (exchange-interaction) processes. Section 4 contains an analysis of surface magnetic currents in a ferrite disk in a cylindrical coordinate system. The currents are described by double-valued functions and one should introduce additional "spinning coordinates" to analyze the current properties. The nature of surface magnetic currents can be understood when analyzing the boundary conditions in a helical coordinate system. Section 5 is devoted to an analysis of MS modes in a ferrite rod in a helical coordinate system. Formally one can distinguish four helical modes in an infinite ferrite rod. However, these helical modes acquire a real physical meaning only in a case of a ferrite disk resonator. In Section 6 we show that there are four basic solutions for helical harmonics in a ferrite disk. One of the main aspects is an analysis of the power flow relations for magnetostatic helical harmonics in a ferrite disk. This analysis is a subject of Section 7. Since helical coordinates are not separable, to get the physically adequate models for MS oscillations in a ferrite disk we have to correlate the obtained results with the ones given from the cylindrical coordinate system. In Section 8 we consider handedness of MS modes in a ferrite disk in a cylindrical coordinate system based on the Dirac coordinates. Section 9 contains conclusive remarks.

## 2. MAGNETIC-DIPOLAR MODES: NEITHER ELECTROMAGNETIC NOR EXCHANGE-INTERACTION WAVES

The fact that magnetic-dipolar waves may have wavelength much smaller than the EM wavelength and, at the same time, much larger than the wavelength of the exchange-interaction waves arises the questions: What are the fields of magnetic dipolar modes, what kind of a spectral problem describes these modes, and what are the excitation problems for such waves? Historically, these questions constituted the subject of numerous discussions (see e.g. [29, 30]). Magnetic-dipolar-mode oscillations in a ferrite sample occupy an intermediate position between two wave processes: the "pure" electromagnetic waves and the exchange-interaction waves. So one might suppose that the light magnons [12] are not subjected to the classical relativistic treatment and, at the same time, the ("real", "heavy") magnon motion laws.

Based on the scalar magnetic-dipolar wave function we are able to obtain the *complete-set functional basis in the energy-eigenstate spectral problem* [11,12]. To get this orthonormal functional basis we attracted neither the notion of the RF electric field, nor the notion of the RF magnetization field. For magnetic-dipolar wave functions there are two *different-nature* limits: the EM-wave limit for small wavenumbers and the exchange-interaction-wave limit for large wavenumbers. Since the MS-wave spectrum forms the complete-set functional basis it can be reduced neither to the EM-wave spectrum, nor to the exchange-interaction-wave spectrum. For magnetic-dipolar-mode oscillations the above limit cases should just only be considered as, respectively, the EM-wave approach and the exchange-interaction-wave approach. We illustrate this statement by consideration of the spectral problems.

### A. A spectral problem for magnetic-dipolar waveguide modes

For MS-wave waveguide modes in ferrite samples the spectral problem was formulated in [11, 12]. The setting of a problem was made for a monochromatic wave process ($\sim e^{i\omega t}$) for two variables: the MS-potential wave function $\psi$ and the magnetic flux density $\vec{B} = -\vec{\vec{\mu}}(\omega) \cdot \nabla \psi$, where $\vec{\vec{\mu}}$ is the permeability tensor. The orthogonality relation is expressed as:

$$(\beta_p - \beta_q) \sum_j \int_{S_j} \left( \hat{R} \widetilde{V}_p^{(j)} \right) \left( \widetilde{V}_q^{(j)} \right)^* dS_j = 0, \tag{4}$$

where

$$\widetilde{V} = \begin{pmatrix} \widetilde{\vec{B}} \\ \widetilde{\varphi} \end{pmatrix}, \tag{5}$$

$$\hat{R} = \begin{pmatrix} 0 & \vec{e}_z \\ -\vec{e}_z & 0 \end{pmatrix}, \tag{6}$$

$\vec{e}_z$ is the unit vector along longitudinal z-axis, $\tilde{\varphi}$ and $\tilde{\vec{B}}$ are MS-wave membrane functions, $S_j$ is a cross section of the *j*-th waveguide layer. The norm of mode $p$ is:

$$N_p = -i\omega \frac{1}{c} \sum_j \int_{S_j} \left(\hat{R}\tilde{\vec{V}}_p^{(j)}\right) \cdot \left(\tilde{\vec{V}}_p^{(j)}\right)^* dS_j = -i\omega \frac{1}{c} \sum_j \int_{S_j} \left[\tilde{\varphi}_p^{(j)} \left(\tilde{\vec{B}}_p^{(j)}\right)^* - \left(\tilde{\varphi}_p^{(j)}\right)^* \tilde{\vec{B}}_p^{(j)} \right] \cdot \vec{e}_z \, dS_j. \qquad (7)$$

Here factor $i\omega \frac{1}{c}$ is used as a dimensional coefficient. For the MS-wave processes, the norm described by Eq. (7) has concrete physical meaning as the power flow density. It becomes clear from the energy balance equation for monochromatic MS waves [11]:

$$\frac{i\omega}{4\pi} \nabla \cdot \left(\psi \vec{B}^* - \psi^* \vec{B}\right) + \frac{i\omega}{4\pi} \left[\vec{B}^* \cdot \vec{\mu}^{-1} \cdot \vec{B} - \vec{B} \cdot \left(\vec{\mu}^*\right)^{-1} \cdot \vec{B}^*\right] = 0. \qquad (8)$$

The first term in the left-hand side (LHS) of Eq. (8) is the divergence of the power flow density and the second term in the LHS of this equation is the density of magnetic losses.

Correct formulation of a spectral problem for the MS-potential wave functions shows that in this case one has the *complete-set energy functional space* [11,12].

**B. Electromagnetic-wave approach**

For MS modes [6, 7] we combine the Landau-Lifshitz torque equation:

$$\frac{\partial \vec{M}}{\partial t} = -\gamma \, \vec{M} \times \vec{H} \qquad (9)$$

(where $\vec{M} = \vec{M}_s + \vec{m}$ is the total magnetization, $\vec{M}_s$ and $\vec{m}$ are the vectors of the saturation and the variable magnetization, respectively) with the equations of magnetostatics:

$$\nabla \times \vec{H} = 0 \qquad (10)$$

and

$$\nabla \cdot \vec{B} = 0. \qquad (11)$$

Combination of Eqs. (9) – (11) implies the fact that $\frac{\partial \vec{B}}{\partial t} \neq 0$. So one can suppose that the entire set of equations contains Eqs. (9) – (11) together with the Faraday-law equation:

$$\nabla \times \vec{E} = -\frac{1}{c} \frac{\partial \vec{B}}{\partial t}. \qquad (12)$$

Taking into account Eq. (10) and Eq. (12) written for a monochromatic wave process, one has:

$$-i\omega\frac{1}{c}\nabla_{\perp}\cdot(\widetilde{\varphi}\vec{\widetilde{B}}^{*}-\widetilde{\varphi}^{*}\vec{\widetilde{B}})=-i\omega\frac{1}{c}(\vec{\widetilde{B}}^{*}\cdot\nabla_{\perp}\widetilde{\varphi}-\vec{\widetilde{B}}\cdot\nabla_{\perp}\widetilde{\varphi}^{*})=$$
$$(\nabla_{\perp}\times\vec{\widetilde{E}}^{*}\cdot\vec{\widetilde{H}}-\nabla_{\perp}\times\vec{\widetilde{E}}\cdot\vec{\widetilde{H}}^{*})=\nabla_{\perp}\cdot(\vec{\widetilde{E}}^{*}\times\vec{\widetilde{H}}+\vec{\widetilde{E}}\times\vec{\widetilde{H}}^{*}),\qquad(13)$$

where tilde means membrane functions and subscript $\perp$ means differentiation over waveguide cross section. So one can express the norm of MS-wave waveguide mode $p$ by the electric- and magnetic-field membrane functions as:

$$N_p = \sum_j \int_{S_j} \left[ \vec{\widetilde{E}}_p^{(j)} \times \left( \vec{\widetilde{H}}_p^{(j)} \right)^* + \left( \vec{\widetilde{E}}_p^{(j)} \right)^* \times \vec{\widetilde{H}}_p^{(j)} \right] \cdot \vec{e}_z \, dS_j . \qquad (14)$$

As it was stated in [7, 31], vector $\frac{c}{4\pi}N_p$ has a physical meaning of the Poynting vector for MS mode $p$. This gives, in particular, a basis for the well-known theories of the MS-wave excitation by an electric current (see e.g. [32, 33]).

Formal mathematical reduction of Eq. (7) to Eq. (14) based on conversion (13), reveals, however, evident physical contradictions. From a classical electromagnetic point of view, one does not have any physical mechanism describing the effect of transformation of the *curl* electric field to the *potential* (quasi-magnetostatic) magnetic field. So for MS modes the norm (14) *does not correspond to the(electromagnetic) Poynting vector* written for curl components of the fields. This underlies the fact that the MS modes are not the electromagnetic waves. Another aspect concerning the non-electromagnetic nature of MS modes was shown recently by McDonald [34]. Since for magnetostatics (in an assumption that ferrite is an isotropic dielectric) $\partial\vec{E}/\partial t=0$, one gets from the Faraday law that $\partial^2\vec{B}/\partial t^2=0$. This is consistent with a magnetic field that *varies linearly with time*: $\vec{B}(\vec{r},t)=\vec{B}_0(\vec{r},t)+\vec{B}_1(\vec{r})t$, where $\vec{B}_0(\vec{r},t)$, $\vec{B}_1(\vec{r})$ are respectively the DC and RF components of the magnetic flux density. The result leads, however, to arbitrary large magnetic fields at early and last times, and is physically excluded. One can conclude that in the spectral problem of the magnetic-dipolar-mode oscillations, the classical Faraday law is inapplicable. It can be used just only for the first-order modes with large MS-wave wavelength.

**C. Exchange-interaction-wave approach**

Maxwell's equations do not imply any wave functions in the magnetostatic approximation. From another point of view, one can suppose that due to the dipole-dipole interaction the microwave components of the magnetization could be described by the wave equations.

The magnetic energy of the thin-film ferrite sample, which includes the energies of the Zeeman interaction with the external magnetic fields (constant bias magnetic field and the time-dependent field of the external excitation signal), dipole-dipole and exchange interactions, can be written in terms of the magnetization-vector amplitudes. Distributions of the magnetization amplitudes are described by the spin-wave modes, which satisfy the *exchange-interaction boundary conditions* (see e.g. [35] and references therein). This variant of solution cannot be acceptable, however, for enough large ferrite samples (with characteristic sizes much exceeding the exchange-interaction spin-wave wavelength). If one supposes to use the magnetization spectra for "thick" ferrite films (films with thickness about tens of micrometers), one should be able to show a consistent formulation of the spectrum problem (a differential operator plus homogeneous boundary conditions) for magnetization modes in such samples.

Attempts to formulate a spectral problem for the magnetization field for "pure" magnetic-dipolar modes were undertaken by many authors. It was stated that these modes are just the natural modes of magnetic dipolar continuum of some definite shape which is immersed in a homogeneous magnetic field. The basis for such a statement the authors found from the fact that for the variable magnetization corresponding to different MS modes in a ferromagnetic spheroid, one can obtain the orthogonality relation. For MS modes the RF magnetization is defined as

$$\vec{m} = -\ddot{\chi}(\omega) \cdot \nabla \psi, \qquad (15)$$

where $\ddot{\chi}$ is the susceptibility tensor. For a ferromagnetic spheroid with the internal DC magnetic field directed along $z$ axis (with use of proper boundary conditions for the MS-potential function and the magnetic flux density), Walker obtained the ortogonality relation for two oscillating MS modes [36]:

$$[\omega^{(\lambda)} + (\omega^{(\nu)})^*] \int_V [\vec{m}_\perp^{(\lambda)} \times (\vec{m}_\perp^{(\nu)})^*] \cdot \vec{e}_z dV = 0, \qquad (16)$$

where $\vec{e}_z$ is the unit vector along $z$ axis and $\vec{m}_\perp$ the transversal-component magnetization vector. There are also other types of such orthogonality relations [37]. All these orthogonality relations were derived, however, from the equations for MS-potential functions but not based on initial formulation of the spectral problem for the magnetization field of MS modes. The spectral picture for the magnetization field takes place only for ferrite samples with characteristic sizes compared to the exchange-interaction spin-wave wavelength. In this case one has the boundary conditions written specifically for the magnetization $\vec{m}$ (the Ament-Rado and Kittel boundary conditions [6]). The above means that there is no complete-spectrum magnetization continuum for MS modes in "thick" films, just as we can see for exchange-interaction spin waves in ultrathin ferromagnetic films.

**D. Excitation problems**

As we showed above, there are no consistent formulations of a spectral problem for "pure" magnetic dipolar modes based on Maxwell's equations or based on an analysis of the magnetization field. Nevertheless, these two limit cases should be considered as certain approaches: the electromagnetic-wave approach – for large MS-wave wavelengths and the exchange-interaction-wave approach – for short MS-wave wavelengths. For the limit cases these approaches can be successfully used in solving the excitation problem.

As we discussed above, the electromagnetic-wave approach gives a basis for the theories of the MS-wave excitation by an electric current [32, 33]. It could be supposed also that in known experiments of excitation of ferrite samples by external RF magnetic fields [15, 16], the electromagnetic-wave approach is applicable for the main (large-wavelength) oscillating MS modes. In this case the coupling between an external RF magnetic field and internal MS wave process takes place because of the MS-mode electric fields found from the Faraday law. The non-homogeneous Maxwell equation is:

$$\nabla \times \vec{E} + i\omega \frac{1}{c} \ddot{\mu} \vec{H} = -j^m. \qquad (17)$$

Here we have the external-source magnetic current appearing due to "surface dipolar magnetic charges", which are induced on opposite sides of a MS-wave waveguide cross section by the external

RF magnetic field $\vec{H}^{ext}$. From Eqs. (10) and (17) one has the excitation equation for an amplitude of MS-wave waveguide mode *p* (see e.g. [38]):

$$\frac{da_p(z)}{dz} + i\beta_p a_p(z) = \frac{1}{N_p} \int_S \vec{j}_\perp^m \cdot \left(\widetilde{\vec{H}}_p^*\right)_\perp dS, \qquad (18)$$

where *S* is a waveguide cross section. Norm $N_p$ is defined from Eq. (14). For a circular cross section, the source $\vec{j}_\perp^m$ is defined as:

$$\vec{j}_\perp^m = i\omega \frac{1}{4\pi} [(\vec{1} - \vec{\mu}) \cdot \vec{H}^{ext}]_\perp, \qquad (19)$$

where $\vec{1}$ is the unit matrix. As a result one obtains:

$$\frac{da_p(z)}{dz} + i\beta_p a_p(z) = \frac{i\omega}{4\pi N_p} \int_S [(\vec{1} - \vec{\mu}) \cdot \vec{H}^{ext}]_\perp \cdot \left(\widetilde{\vec{H}}_p^*\right)_\perp dS. \qquad (20)$$

A ferrite disk can be considered as a MS-wave waveguide section. So based on Eq. (20) one can solve the excitation problem of ferrite samples by external RF magnetic fields. The excitation integral in Eq. (20) differs from the excitation integral obtained in a frame of the exchange-interaction-wave approach. In the last case the excitation integral written for magnetization mode *n* in a ferrite sample is defined as $\int_V \vec{H}^{ext} \cdot \vec{m}_n^* dV$ [1, 6].

## 3. THE FARADAY-LIKE EFFECT FOR MAGNETOSTATIC MODES IN AN AXIALLY MAGNETIZED FERRITE ROD

The above analysis shows that the complete-spectrum MS oscillations differ from spectra of electromagnetic waves and exchange-interaction waves. Another important distinction of MS waves from electromagnetic waves can be illustrated from consideration of the Faraday effect.
 When a plane polarized electromagnetic wave is incident normally on a ferromagnet whose magnetization axis lies along the normal, there will be two normal modes in a ferrite. The wavenumbers of these modes are different due to the off-diagonal terms of the permeability tensor:

$$k_\pm = \frac{\omega}{c}\sqrt{\varepsilon}\sqrt{\mu \pm \mu_a}, \qquad (21)$$

where $\mu$ and $\mu_a$ are diagonal and off-diagonal components of the permeability tensor. Since for electromagnetic waves in ferromagnets there are two values of the wave vector to a given frequency, it follows that the rotation of the plane of polarization (the Faraday effect) is possible in ferromagnets [1, 6]. When the wave propagation changes to the opposite direction, the rotation angle will be the same after the wave reflection. At the same time, in a point where magnetization becomes oppositely directed, the rotation angle changes its sign [1, 6]. For electromagnetic-wave ferrite cylinder resonator one has different resonant conditions (and, therefore, different resonant frequencies) for modes with $k_+$ and $k_-$.

For MS modes the Faraday effect is displayed by another way than for electromagnetic waves in gyrotropic media. Let us consider a ferrite rod with radius $\Re$ axially magnetized along $z$-axis. In a cylindrical coordinate system $(r, \theta, z)$ the Walker equation (1) has a form:

$$\mu \left( \frac{\partial^2 \psi}{\partial r^2} + \frac{1}{r} \frac{\partial \psi}{\partial r} + \frac{1}{r^2} \frac{\partial^2 \psi}{\partial \theta^2} \right) + \frac{\partial^2 \psi}{\partial z^2} = 0. \tag{22}$$

There is no off-diagonal term $\mu_a$ in this equation. However, the off-diagonal term emerges in the boundary conditions. As a result of solution of a boundary problem (supposing that $\psi \sim e^{-i(\nu\theta + \beta z)}$), one has the following characteristic equation [19]:

$$(-\mu)^{\frac{1}{2}} \left( \frac{J_\nu'}{J_\nu} \right)_{r=\Re} + \left( \frac{K_\nu'}{K_\nu} \right)_{r=\Re} - \frac{\mu_a \nu}{|\beta|\Re} = 0, \tag{23}$$

where $J_\nu, J_\nu', K_\nu,$ and $K_\nu'$ are the Bessel functions of integer order $\nu$ for real and imaginary arguments and their derivatives with respect to the argument.

It is evident from Eq. (23) that for different signs of $\nu$ we have different solutions for $\beta$. In other words, the left-hand-rotation and the right-hand-rotation waves are non-degenerate with respect to the wave number. This underlies the physics of the Faraday-like effect for MS modes in an axially magnetized ferrite rod. As we can see from Eq. (23), the sign of $\beta$ does not influence on the dispersion relation. So after the wave reflection the rotation will have the same direction. In a normally magnetized ferrite disk resonator one will have different resonance frequencies for MS modes with different signs of $\nu$.

For MS modes propagating along a ferrite rod with a wave number $\beta$, we rewrite Eq. (22) in an operator form:

$$\hat{G}_\perp \psi - \beta^2 \psi = 0, \tag{24}$$

where

$$\hat{G}_\perp \equiv \mu \nabla_\perp^2, \tag{25}$$

$\nabla_\perp^2$ is the two-dimensional Laplace operator. For propagating MS modes, operator $\hat{G}_\perp$ is the positive definite operator. A double integration by parts (the Green theorem) on $S$ – a square of a cross section of an open MS-wave waveguide – of the integral $\int_S (\hat{G}_\perp \psi) \psi^* dS$, gives the following boundary condition for the *energy* orthonormality [12]:

$$\mu \left( \frac{\partial \psi}{\partial r} \right)_{r=\Re^-} - \left( \frac{\partial \psi}{\partial r} \right)_{r=\Re^+} = 0 \tag{26}$$

or

$$\mu (H_r)_{r=\Re^-} - (H_r)_{r=\Re^+} = 0, \tag{27}$$

where $H_r = -\dfrac{\partial \psi}{\partial r}$ is a radial component of the RF magnetic field, $\Re^-$ and $\Re^+$ designate, respectively, the inner (ferrite) and outer (dielectric) regions of a MS-wave waveguide. The orthogonality relation for MS modes $p$ and $q$ is expressed as:

$$\left(\beta_p^2 - \beta_q^2\right)\int_S \psi_p \psi_q^* dS = 0. \qquad (28)$$

One can see that two MS modes, which are distinguished only by a sign of $\nu$, are degenerate with respect to the MS energy.

For operator $\hat{G}_\perp$, the boundary condition of the MS-potential continuity together with the boundary condition (26) [or (27)] are the so-called *essential* boundary conditions. Such boundary conditions are well known for different mathematical physics problems (see e.g. [39]). When the essential boundary conditions are used, the MS-potential eigen functions of operator $\hat{G}_\perp$ form a *complete basis in an energy functional space*, and the functional describing an average quantity of energy, has a minimum at the energy eigenfunctions [39]. In [12], calculations of complete-set energy spectra of MS oscillations in a ferrite disk resonator were made based on the essential boundary conditions.

The essential boundary conditions differ from the homogeneous electrodynamics boundary conditions (or *natural* boundary conditions [39]) at $r = \Re$. The natural boundary conditions demand continuity for the radial component of the magnetic flux density:

$$\mu(H_r)_{r=\Re^-} - (H_r)_{r=\Re^+} = -i\mu_a (H_\theta)_{r=\Re^-}, \qquad (29)$$

where $H_\theta = -\dfrac{1}{r}\dfrac{\partial \psi}{\partial \theta}$ is an azimuth component of the RF magnetic field. Supposing that $\psi \sim e^{-i\nu\theta}$ one can rewrite (29) as

$$\mu\left(\dfrac{\partial \psi}{\partial r}\right)_{r=\Re^-} - \left(\dfrac{\partial \psi}{\partial r}\right)_{r=\Re^+} = -\dfrac{\mu_a}{\Re}\nu(\psi)_{r=\Re^-}. \qquad (30)$$

The main feature of the natural boundary conditions (in comparison with the essential boundary conditions) arises from the quantity of annular magnetic field $(H_\theta)_{r=\Re^-}$. One can see that this is a *singular* field, which exists only in an infinitesimally narrow cylindrical layer abutting (from a ferrite side) to the ferrite-dielectric border. One does not have any special conditions connecting radial and azimuth components of magnetic fields on other (inner or outer) circular contours, except contour $L = 2\pi\Re$. Characteristic equation (23) was derived based on natural boundary conditions (29) [or (30)]. It shows that because of annular magnetic field $(H_\theta)_{r=\Re^-}$ one has an *additional degree of freedom*. This degree of freedom gives two – the left-handed-rotation and the right-handed-rotation – solutions, *which are degenerate energetically*.

Let us introduce formally the notion of an effective circular magnetic current:

$$\vec{j}^{\,m}(z) \equiv \dfrac{1}{4\pi}i\omega\mu_a \vec{H}_\theta(z). \qquad (31)$$

We can rewrite the boundary condition (29) as follows:

$$\delta(r-\Re)\left[\dfrac{1}{4\pi}\omega\mu(H_r)_{r=\Re^-} - \dfrac{1}{4\pi}\omega(H_r)_{r=\Re^+}\right] = -i^m, \qquad (32)$$

where $i^m$ is a density of an *effective surface magnetic current* defined as

$$\vec{i}^{\,m}(z) \equiv \delta(r-\Re)\frac{1}{4\pi}i\omega\mu_a(\vec{H}_\theta(z))_{r=\Re^-} = \delta(r-\Re)\vec{j}^{\,m}(z). \qquad (33)$$

In fact, the Faraday-like effect (the Faraday rotation) in an axially magnetized MS-wave ferrite rod is due to the surface magnetic current. In a ferrite cylinder, this current cannot perturb the energy functional space. This differs from the case of a flat ferrite disk. As we will show below, in a normally magnetized disk a circular surface magnetic current causes an electric moment of a sample. So one can suppose that an external normal RF electric field may affect on the energy functional space of MS oscillations and, therefore, the oscillating spectrum will be perturbed. In paper [40], it is shown experimentally that an external RF electric field really perturbs magnetic-dipolar oscillations in a ferrite disk.

A circular surface magnetic current is a singular current. Certainly, the standard problems with cylindrical symmetry solutions – the cylindrical (Bessel) functions – should not be dependent on a sign of an azimuth variation. To understand more clearly why such a singularity appears, let us consider the MS-wave solutions for a general case of an axially magnetized ferrite rod with the permeability-tensor components dependent on a radial coordinate: $\vec{\mu} = \vec{\mu}(r)$. In this case we have the following differential equation for the MS-potential function:

$$\nabla \cdot (\vec{\mu}(r) \cdot \nabla \psi) = 0. \qquad (34)$$

It is not difficult to show that in a cylindrical coordinate system this equation has a form:

$$\mu\left(\frac{\partial^2\psi}{\partial r^2} + \frac{1}{r}\frac{\partial\psi}{\partial r} + \frac{1}{r^2}\frac{\partial^2\psi}{\partial \theta^2}\right) + \frac{\partial\mu}{\partial r}\frac{\partial\psi}{\partial r} + i\frac{1}{r}\frac{\partial\mu_a}{\partial r}\frac{\partial\psi}{\partial \theta} + \frac{\partial^2\psi}{\partial z^2} = 0. \qquad (35)$$

One can see that in this equation separation of variables is impossible and therefore an analytical solution cannot be found. Even assuming possible numerical solutions, it becomes clear, however, that (because of the presence of the azimuth-first-derivative term: $\frac{\partial\psi}{\partial\theta}$) these solutions will not be described by single-valued functions. In a supposition that a ferrite rod has homogeneous cross-section parameters (and, therefore, can be characterized by Eq. (22)), on a lateral cylindrical surface one has a sharp transition of the permeability-tensor components. In particular, there is a sharp transition of the off-diagonal component $\mu_a$ (from $\mu_a$ to zero). So in a boundary region one has non-single-valued functions. As we will show below, the boundary non-singlevaluedness can be avoided for MS-wave oscillations in a ferrite disk in a helical coordinate system.

## 4. SURFACE MAGNETIC CURRENT IN A NORMALLY MAGNETIZED FERRITE DISK

In a normally magnetized ferrite-disk resonator with a small thickness to diameter ratio, separation of variables is possible. In this case, the MS potential $\psi$ is represented in a cylindrical coordinate system as [11]:

$$\psi = \sum_{m,n} A_{m,n}\tilde{\xi}_m(z)\tilde{\varphi}_n(r,\theta), \qquad (36)$$

where $A_{m,n}$ is a MS mode amplitude, $\tilde{\xi}_m(z)$ and $\tilde{\varphi}_n(r,\theta)$ are dimensionless functions describing, respectively, $m$ "thickness" ($z$ coordinate) and $n$ "in-plane", or "flat" (radial $\rho$ and azimuth $\alpha$ coordinates) MS modes. In a ferrite disk with a small thickness/diameter ratio, the spectrum of "thickness modes" is very "rare" compared to the "dense" spectrum of "flat" modes. The spectrum of "flat" modes is completely included into the wave-number region of a fundamental "thickness" mode [12, 20]. It means that the spectral properties of a resonator can be entirely described based on consideration of only a fundamental "thickness" mode. These spectral properties are characterized by the energy eigenstates. In [11], an analysis of the energy eigenstates is made for quasimonochromatic processes. In other words, we are dealing with a slowly varying envelope function. The complete envelope function is a plane wave (with a cylindrical cross-sectional area and a certain wave vector) multiplied by a slowly varying amplitude function. This amplitude function describes the confined motion of quasiparticles – the light magnons [12] – along the z axis. MS-potential functions $\tilde{\varphi}$ are functions with finite energy. For these functions, the main feature of the natural boundary conditions (29) and (30) arises from the quantity of the azimuth magnetic field. As we discussed above, this is a singular field, which exists only in an infinitesimally narrow cylindrical layer abutting (from a ferrite side) to the ferrite-dielectric border. In accordance with a first-order differential equation (30), the functions $\tilde{\varphi}$ are dependent on a sign of $\nu$. So one can distinguish the "right" and the "left" functions $\tilde{\varphi}$ and thus functions $\tilde{\varphi}$ cannot be considered as single-valued functions.

Let us consider a circulation of vector $\vec{H}$ along a circular contour $L = 2\pi\Re$, where $\Re$ is a disk radius. Since on this contour $H_\theta^{(L)} = -\frac{1}{\Re}\frac{\partial \tilde{\varphi}^{(L)}}{\partial \theta}$, we can write the circulation as $C = \nu \int_0^{2\pi} \tilde{\varphi}^{(L)} d\theta$. For a single-valued function $\tilde{\varphi}$ circulation $C$ should be equal to zero. This fact follows also from the MS description ($\nabla \times \vec{H} = 0$) of a thin ferrite disk. Our analysis shows, however, that this circulation has a non-zero quantity. The solution depends on both a modulus and a sign of $\nu$. We have a sequence of angular eigenvalues restricted from above and below by values equal in a modulus and different in a sign, which we denote as $\pm s^e$. The difference $2s^e$ between the largest and smallest values is an integer or zero. So $s^e$ can have values $0, \pm 1/2, \pm 1, \pm 3/2,..., \nu/2$. At a full-angle "in-plane" rotation (at an angle equal to $2\pi$) of a system of coordinates, the "flat" functions $\tilde{\varphi}$ with integer values $s^e$ return to their initial states (single-valued functions) and the "flat" functions $\tilde{\varphi}$ with the half-integer values $s^e$ will have an opposite sign (double-valued functions). The only possibility in our case is to suggest that $s^e$ are the half-integer quantities. Because of the *double-valuedness properties* of MS-potential functions on a lateral surface of a ferrite disk resonator, we can talk about the "spinning-type rotation" along a border contour $L$. Along with the well-known notion of the "*magnetic spin*" as a quantity correlated with the eigen magnetic moment of a particle, we introduce (for the phenomena under consideration) the notion of the "*electric spin*" as a quantity correlated with the eigen electric moment. In fact, this vector should be viewed as describing a "pseudospin", in analogy to the two-component spinor describing the particle's real (physical) spin. For integer quantities $s^e$ the eigen electric moment is equal to zero, but it is non-zero for half-integer values $s^e$. This makes clear that in the above consideration superscript *e* in $s^e$ means "electric".

Let $\tilde{\varphi}_+$ and $\tilde{\varphi}_-$ be, respectively, the right-hand and the left-hand side circularly polarized "flat" MS functions. Similarly to the three $2 \times 2$ Pauli matrices, we represent the matrices characterizing the components of the "electric spin" as:

$$\hat{s}_+^e = \begin{pmatrix} 0 & 0 \\ 1 & 0 \end{pmatrix}, \qquad \hat{s}_-^e = \begin{pmatrix} 0 & 1 \\ 0 & 0 \end{pmatrix}, \qquad \hat{s}_z^e = |u|\begin{pmatrix} 1 & 0 \\ 0 & -1 \end{pmatrix}. \tag{37}$$

Quantity *u* characterizes the "spin coordinates". To distinguish the "right" and "left" MS-potential functions we should write that

$$u = k\frac{1}{2}, \qquad (38)$$

where *k* is an integer odd (positive or negative) quantity.

One should suppose that to have the system stability the "spin states", characterizing by quantities *u*, ought to be in a certain *synchronism* (in other words, to be *correlated*) with the "orbit states", characterizing by quantities $v$. The mutual synchronism can be expressed as:

$$u = \frac{v}{n}, \qquad (39)$$

where *n* is a positive integer (odd or even) quantity. The last equation means that the angular velocities of the "spin-state" variations are *n* times more than the angular velocities of the "orbit-state" variations.

Connection between Eqs. (38) and (39) gives:

$$k = \frac{2v}{n}. \qquad (40)$$

This leads to the following conclusion:

$$k = \pm 1 \quad \text{with} \quad n = 2|v| \qquad (41)$$

for any even number $v$, and

$$k = \pm 1 \text{ and } k = \pm|v| \quad \text{with} \quad n = 2|v| \text{ and } n = 2 \qquad (42)$$

for odd numbers $v$. As examples, we can write:

$$\begin{aligned}
&\text{for } v = \pm 1 \quad \rightarrow \quad u = \pm\frac{1}{2}; \\
&\text{for } v = \pm 2 \quad \rightarrow \quad u = \pm\frac{1}{2}; \\
&\text{for } v = \pm 3 \quad \rightarrow \quad u = \pm\frac{1}{2}, \pm\frac{3}{2}; \\
&\text{for } v = \pm 4 \quad \rightarrow \quad u = \pm\frac{1}{2}; \\
&\text{for } v = \pm 5 \quad \rightarrow \quad u = \pm\frac{1}{2}, \pm\frac{5}{2}; \\
&\text{etc.}
\end{aligned} \qquad (43)$$

These quantities of *u* show which spin coordinates are relevant in the system. It can be shown [41] that operators (37) do not change their signs under the space reflection operation. So operators $\hat{s}_+^e$, $\hat{s}_-^e$ and $\hat{s}_z^e$ can be considered as components of an *axial vector*.

For monochromatic time-dependent process (any quantity is $\sim e^{i\omega t}$) we consider now the "border" MS-potential "flat" functions $\widetilde{\widetilde{\varphi}}$ on contour *L*. There are singular functions describing the "spin states". For *k*-th "border" eigenfunction we can write:

$$\widetilde{\widetilde{\varphi}}_k = B_k e^{-iu_k\theta}, \tag{44}$$

where $B_k$ is an amplitude coefficient. Introducing function $\widetilde{\widetilde{\varphi}}$, we have to note that this is not an independent function with respect to function $\widetilde{\varphi}$, but the function showing certain *additional properties, additional states* of the MS-potential scalar wave function. For a certain "thickness" mode and a certain "flat" mode we can represent the $\theta$-component of the "border" (singular) magnetic field as:

$$H_\theta^{(L)}(z) = -A\widetilde{\xi}(z)\nabla_\theta \widetilde{\widetilde{\varphi}} = -A\widetilde{\xi}(z)\frac{1}{\Re}\frac{\partial \widetilde{\widetilde{\varphi}}}{\partial \theta}\bigg|_{r=\Re^-}. \tag{45}$$

For a circular effective boundary magnetic current we have now [see Eq. (33)]:

$$\left(i^m(z)\right)_k = -A\widetilde{\xi}(z)\frac{i\omega\mu_a}{4\pi\Re}\frac{\partial \widetilde{\widetilde{\varphi}}}{\partial \theta}\bigg|_{r=\Re^-} = -A\widetilde{\xi}(z)\frac{\omega\mu_a}{4\pi\Re}u_k B_k e^{-iu_k\theta}. \tag{46}$$

The circular surface magnetic current does not exist due to only precession of magnetization. It appears because of the *combined effect* of precession in a ferrite material and "spinning rotation" caused by the special-type boundary conditions.

Circulation of current $i^m$ along contour $L$ gives a nonzero quantity when $u_k$ is a number divisible by $\frac{1}{2}$:

$$D_k(z) = \oint_L (i^m)_k \, dl = \Re \int_0^{2\pi} (i^m)_k \, d\theta = iA\widetilde{\xi}(z)\frac{\omega\mu_a}{2\pi}B_k. \tag{47}$$

Since circulation $D_k(z)$ is a non-zero quantity, one can define an electric moment of a whole ferrite disk resonator (in a region far away from a disk) as

$$a_k^e = -i\frac{1}{2c}\int_0^d dz \oint_{L=2\pi\Re}(\vec{r}\times\vec{i}^m)\cdot\vec{e}_z \, dl = A\frac{\omega\mu_a}{4\pi c}\Re B_k \int_0^d \widetilde{\xi}(z)dz. \tag{48}$$

The off-diagonal component of the permeability tensor $\mu_a$ can be correlated with a *magnetic vector of gyration* [6]:

$$\vec{g}^m = \frac{\mu_a}{4\pi}\vec{e}_z, \tag{49}$$

where $\vec{e}_z$ is the unit vector along *z*-axis. A sign of $g^m$ corresponds to a sign of $\mu_a$. A sign of amplitude $B_k$ depends on orientation of vector $\vec{s}^e$ ($\vec{s}^e = s^e \vec{e}_z$) with respect to *z*-axis. One can distinguish two cases: $\vec{s}^e \cdot \vec{g}^m > 0$ and $\vec{s}^e \cdot \vec{g}^m < 0$. So (for a given direction of a normal bias magnetic field) the MS-wave wavefunction is a four-component wave function, or a bispinor. The *property of helicity* (a spin orientation is not separated from an orientation of a linear momentum) is well-known in elementary particle physics. In our case, a spin orientation $\vec{s}^e$ is not separated from an orientation of a linear momentum $\vec{a}^e$, but *taking into account also orientation of vector $\vec{g}^m$* [22].

Surface magnetic currents are eigencurrents. The nature of these currents can be understood in a helical coordinate system. In a helical coordinate system the azimuth number does not necessarily have to be an integer as it would in a standard cylindrical system [23-25]. This gives an explanation why the "spinning rotation" on boundary contour *L* takes place. When a dynamical process for the right-hand helical wave is described in a right-handed coordinate system, a dynamical process for the left-hand helical wave propagation should be described in a left-handed coordinate system. In the left-handed coordinate system a sign of $\mu_a$ may be opposite to that one has in the right-handed coordinate system. This becomes clear from consideration of the Landau-Lifshitz equation for harmonic RF magnetization $\vec{m}$ [6]:

$$i\omega \vec{m} + \gamma \, \vec{m} \times \vec{H}_0 = -\gamma \, \vec{M}_0 \times \vec{h} \,. \tag{50}$$

For a given directions of bias magnetic field $\vec{H}_0$ and saturation magnetization $\vec{M}_0$, and for given RF magnetic field $\vec{h}$ and RF magnetization $\vec{m}$ one has opposite signs for vector products in the right-handed and left-handed coordinate systems. This shows that $\mu_a$ has different signs in the right-handed and left-handed coordinate systems.

The helical surface magnetic current $i^m$ takes place due to a combined effect of a circular and axial motion processes. The upper ($z = d$) and lower ($z = 0$) points are the *turn points*. If in a turn point an interference of two-type helical harmonics takes place, the surface magnetic current circumscribes a *double helix*. Let a right-hand helical wave $\widetilde{\widetilde{\varphi}}$ propagates from point $z = 0$ to point $z = d$. After point $z = d$ the wave $\widetilde{\widetilde{\varphi}}$ can propagate to point $z = 0$ along a left-hand helix. For a disk with a small thickness-to-diameter ratio it can be assumed that in a turn point a surface magnetic current $i^m$ is expressed as [see Eq. (46)]:

$$i^m \sim \mu_a \frac{\partial \widetilde{\widetilde{\varphi}}}{\partial \theta} \,. \tag{51}$$

If in a turn point the quantities $\mu_a$ and $\frac{\partial \widetilde{\widetilde{\varphi}}}{\partial \theta}$ remain their signs, a sign of $i^m$ will also preserve its sign. There is a necessary condition for surface magnetic current to circumscribe a double helix. These statements become clearer when we analyze MS modes in a helical coordinate system.

## 5. CHARACTERISTIC EQUATIONS FOR HELICAL MS WAVES IN A FERRITE ROD

It is appropriate to start an analysis of the MS-wave problem in a helical coordinate system from consideration of an infinite axially magnetized ferrite rod.

For a ferrite magnetized along *z* axis the permeability tensor has a form [6]:

$$\vec{\vec{\mu}} = \begin{bmatrix} \mu & i\mu_a & 0 \\ -i\mu_a & \mu & 0 \\ 0 & 0 & 1 \end{bmatrix}. \tag{52}$$

In a cylindrical coordinate system, the magnetic flux density components are correlated with the magnetic field components as:

$$B_r = \mu H_r + i\mu_a H_\theta,$$
$$B_\theta = \mu H_\theta - i\mu_a H_r, \qquad (53)$$
$$B_z = H_z.$$

Taking into account the vector relations for cylindrical $(r,\theta,z)$ and helical $(r,\phi,\zeta)$ coordinates [23], we have from (53) the following expressions for components in a helical coordinate system:

$$B_r = i\mu_a H_\phi \cos\alpha_0 + \mu H_r,$$
$$B_\phi = \mu H_\phi - i\frac{\mu_a}{\cos\alpha_0} H_r, \qquad (54)$$
$$B_\zeta = H_\zeta + (1-\mu)H_\phi \sin\alpha_0 + i\mu_a H_r \tan\alpha_0.$$

In cylindrical coordinates a magnetic field is expressed by magnetostatic potential $\psi$ as:

$$H_r = -(\nabla\psi)_r = -\frac{\partial\psi}{\partial r},$$
$$H_\theta = -(\nabla\psi)_\theta = -\frac{1}{r}\frac{\partial\psi}{\partial\theta}, \qquad (55)$$
$$H_z = -(\nabla\psi)_z = -\frac{\partial\psi}{\partial z}.$$

With use of transformations for differential relations [23] we obtain from (55) for components of a magnetic field in helical coordinates:

$$H_r = -(\nabla\psi)_r = -\frac{\partial\psi}{\partial r},$$
$$H_\phi = -(\nabla\psi)_\phi = -\frac{1}{r\cos\alpha_0}\frac{\partial\psi}{\partial\theta} = -\frac{1}{\cos\alpha_0}\left(\frac{1}{r}\frac{\partial\psi}{\partial\phi} - \tan\alpha_0 \frac{\partial\psi}{\partial\zeta}\right), \qquad (56)$$
$$H_\zeta = -(\nabla\psi)_\zeta = -\frac{\partial\psi}{\partial z} + \frac{\tan\alpha_0}{r}\frac{\partial\psi}{\partial\theta} = \frac{\tan\alpha_0}{r}\frac{\partial\psi}{\partial\phi} - \frac{1}{\cos^2\alpha_0}\frac{\partial\psi}{\partial\zeta}.$$

Based on Eqs. (54) and (56) we have for components of a magnetic flux density:

$$B_r = -\left[\mu\frac{\partial\psi}{\partial r} + i\mu_a\left(\frac{1}{r}\frac{\partial\psi}{\partial\phi} - \tan\alpha_0\frac{\partial\psi}{\partial\zeta}\right)\right],$$
$$B_\phi = -\frac{1}{\cos\alpha_0}\left[\mu\left(\frac{1}{r}\frac{\partial\psi}{\partial\phi} - \tan\alpha_0\frac{\partial\psi}{\partial\zeta}\right) - i\mu_a\frac{\partial\psi}{\partial r}\right], \qquad (57)$$
$$B_\zeta = -\tan\alpha_0\left[\frac{2}{\sin 2\alpha_0}\frac{\partial\psi}{\partial\zeta} - \frac{1}{r}\frac{\partial\psi}{\partial\phi} + (1-\mu)\left(\frac{1}{r}\frac{\partial\psi}{\partial\phi} - \tan\alpha_0\frac{\partial\psi}{\partial\zeta}\right) + i\mu_a\frac{\partial\psi}{\partial r}\right].$$

With the use of Waldron's equation for the divergence [23] we have:

$$\nabla\cdot\vec{B} = \frac{1}{r}\frac{\partial}{\partial r}(rB_r) + \frac{\cos\alpha_0}{r}\frac{\partial B_\phi}{\partial\phi} + \frac{\partial B_\zeta}{\partial\zeta} = 0. \qquad (58)$$

Based on Eqs. (57) and (58) we obtain after some transformations the Walker equation in helical coordinates:

$$\frac{\partial^2 \psi}{\partial r^2} + \frac{1}{r}\frac{\partial \psi}{\partial r} + \frac{1}{r^2}\frac{\partial^2 \psi}{\partial \phi^2} + \left(\frac{1}{\mu} + \tan^2 \alpha_0\right)\frac{\partial^2 \psi}{\partial \zeta^2} - 2\frac{1}{r}\tan \alpha_0 \frac{\partial^2 \psi}{\partial \phi \partial \zeta} = 0. \qquad (59)$$

Outside a ferrite region (where $\mu = 1$) we have the Laplace equation in helical coordinates [23, 25]:

$$\frac{\partial^2 \psi}{\partial r^2} + \frac{1}{r}\frac{\partial \psi}{\partial r} + \frac{1}{r^2}\frac{\partial^2 \psi}{\partial \phi^2} + \left(1 + \tan^2 \alpha_0\right)\frac{\partial^2 \psi}{\partial \zeta^2} - 2\frac{1}{r}\tan \alpha_0 \frac{\partial^2 \psi}{\partial \phi \partial \zeta} = 0. \qquad (60)$$

Following Overfelt's approach [25], we assume that solutions of Eqns. (59) and (60) are found as:

$$\psi(r, \phi, \zeta) = R(r)P(\phi)Z(\zeta). \qquad (61)$$

By substituting Eq. (61) into Eq. (59), we obtain:

$$\frac{R''(r)}{R(r)} + \frac{R'(r)}{rR(r)} + \frac{P''(\phi)}{r^2 P(\phi)} + \left(\frac{1}{\mu} + \tan^2 \alpha_0\right)\frac{Z''(\zeta)}{Z(\zeta)} - 2\frac{1}{r}\tan \alpha_0 \frac{P'(\phi)Z'(\zeta)}{P(\phi)Z(\zeta)} = 0. \qquad (62)$$

Similarly, we have from Eqs. (60) and (61):

$$\frac{R''(r)}{R(r)} + \frac{R'(r)}{rR(r)} + \frac{P''(\phi)}{r^2 P(\phi)} + \left(1 + \tan^2 \alpha_0\right)\frac{Z''(\zeta)}{Z(\zeta)} - 2\frac{1}{r}\tan \alpha_0 \frac{P'(\phi)Z'(\zeta)}{P(\phi)Z(\zeta)} = 0. \qquad (63)$$

In Eqs. (62) and (63) primes denote the corresponding differentiation.

Eqs. (62) and (63) can be reduced to an ordinary differential equation in the variable $r$ alone, being not separated, however, with respect to the $\phi$ and $\zeta$ coordinates. Like Overfelt's approach [25], let us find solutions of Eqs. (62) and (63) in supposition that $P(\phi)$ and $Z(\zeta)$ are exponential functions:

$$\begin{aligned} P(\phi) &= A\exp(\pm \Phi \phi), \\ Z(\zeta) &= B\exp(\pm \Theta \zeta). \end{aligned} \qquad (64)$$

In a general case, $\Phi$ and $\Theta$ are complex quantities:

$$\begin{aligned} \Phi &= \lambda + iw, \\ \Theta &= \alpha + i\beta. \end{aligned} \qquad (65)$$

Here the constants $\lambda, w, \alpha,$ and $\beta$ are all assumed to be real and positive.

Upon substituting Eq. (64) into Eqs. (62) and (63) we obtain, respectively,

$$\frac{R''(r)}{R(r)} + \frac{R'(r)}{rR(r)} + \frac{1}{\mu}\Theta^2 + \frac{1}{r^2}\left[\Phi^2 + (\overline{p}\Theta)^2 - 2\overline{p}(\pm \Phi)(\pm \Theta)\right] = 0 \qquad (66)$$

and

$$\frac{R''(r)}{R(r)}+\frac{R'(r)}{rR(r)}+\Theta^2+\frac{1}{r^2}\left[\Phi^2+(\bar{p}\Theta)^2-2\bar{p}(\pm\Phi)(\pm\Theta)\right]=0. \tag{67}$$

Here $\bar{p}\equiv p/2\pi$.

Every of these equations may have physical meaning for two cases: (a) both exponents in (64) have the same signs, positive or negative, and (b) both exponents in (64) have different signs. We, conventionally, will call the first type of the above equations as the (‡) case and the second type – as the (±) case. For the (‡) case we rewrite Eqs. (66) and (67) for the *r*-dependent part of the MS-potential function as:

$$\frac{\partial^2\psi(r)}{\partial r^2}+\frac{1}{r}\frac{\partial\psi(r)}{\partial r}+\left[\frac{1}{\mu}\Theta^2+\frac{1}{r^2}(\Phi-\bar{p}\Theta)^2\right]\psi(r)=0 \tag{68}$$

and

$$\frac{\partial^2\psi(r)}{\partial r^2}+\frac{1}{r}\frac{\partial\psi(r)}{\partial r}+\left[\Theta^2+\frac{1}{r^2}(\Phi-\bar{p}\Theta)^2\right]\psi(r)=0. \tag{69}$$

For the (±) case we have for ferrite and dielectric regions, respectively:

$$\frac{\partial^2\psi(r)}{\partial r^2}+\frac{1}{r}\frac{\partial\psi(r)}{\partial r}+\left[\frac{1}{\mu}\Theta^2+\frac{1}{r^2}(\Phi+\bar{p}\Theta)^2\right]\psi(r)=0 \tag{70}$$

and

$$\frac{\partial^2\psi(r)}{\partial r^2}+\frac{1}{r}\frac{\partial\psi(r)}{\partial r}+\left[\Theta^2+\frac{1}{r^2}(\Phi+\bar{p}\Theta)^2\right]\psi(r)=0. \tag{71}$$

The general solutions of Eqs. (68) – (71) are expressed by Bessel functions of complex arguments and complex orders. These analytic solutions are referred as the *helical Bessel functions* [25]. In our analysis of helical MS waves we will be interested in only the cases of real arguments and orders assuming no attenuation or gain in a system. So we have $\Theta=i\beta$ and $\Phi=iw$.

For the (‡) helical MS wave propagating along an infinite ferrite rod we have from Eqs. (68) and (69):

$$\frac{\partial^2\psi(r)}{\partial r^2}+\frac{1}{r}\frac{\partial\psi(r)}{\partial r}-\left[\frac{\beta^2}{\mu}+\frac{1}{r^2}(w-\bar{p}\beta)^2\right]\psi(r)=0 \tag{72}$$

inside a ferrite rod $(r\leq\Re)$ and

$$\frac{\partial^2\psi(r)}{\partial r^2}+\frac{1}{r}\frac{\partial\psi(r)}{\partial r}-\left[\beta^2+\frac{1}{r^2}(w-\bar{p}\beta)^2\right]\psi(r)=0 \tag{73}$$

outside a ferrite rod $(r\geq\Re)$. A physically acceptable solution for Eq. (72) is possible only for a negative quantity $\mu$. This solution is expressed by Bessel function of a real argument:

$$(\psi(r))_{r<\Re} = c_1 J_{(w-\bar{p}\beta)}[\beta(-\mu)^{1/2} r]. \tag{74}$$

A solution of Eq. (73) is expressed by Bessel function of an imaginary argument:

$$(\psi(r))_{r>\Re} = d_1 K_{(w-\bar{p}\beta)}(\beta r). \tag{75}$$

For the (±) helical MS wave we have from Eqs. (70) and (71):

$$\frac{\partial^2 \psi(r)}{\partial r^2} + \frac{1}{r}\frac{\partial \psi(r)}{\partial r} - \left[\frac{\beta^2}{\mu} + \frac{1}{r^2}(w+\bar{p}\beta)^2\right]\psi(r) = 0 \tag{76}$$

inside a ferrite rod $(r \leq \Re)$ and

$$\frac{\partial^2 \psi(r)}{\partial r^2} + \frac{1}{r}\frac{\partial \psi(r)}{\partial r} - \left[\beta^2 + \frac{1}{r^2}(w+\bar{p}\beta)^2\right]\psi(r) = 0 \tag{77}$$

outside a ferrite rod $(r \geq \Re)$. Similarly to Eqs. (74) and (75), solutions of Eqs. (76) and (77) are expressed as:

$$(\psi(r))_{r<\Re} = c_2 J_{(w+\bar{p}\beta)}[\beta(-\mu)^{1/2} r] \tag{78}$$

inside a ferrite rod $(r \leq \Re)$ and

$$(\psi(r))_{r>\Re} = d_2 K_{(w+\bar{p}\beta)}(\beta r) \tag{79}$$

outside a ferrite rod $(r \geq \Re)$. Coefficients $c_{1,2}, d_{1,2}$ in Eqs. (74), (75), (78), and (79) are amplitude coefficients.

On a cylindrical surface of a ferrite rod we have the boundary conditions:

$$(\psi)_{r=\Re^-} = (\psi)_{r=\Re^+} \tag{80}$$

and

$$(B_r)_{r=\Re^-} = (B_r)_{r=\Re^+}. \tag{81}$$

Based on the first equation (54), the last equation can be rewritten as:

$$\left(i\mu_a H_\phi \frac{\Re}{\sqrt{\Re^2 + \bar{p}^2}} + \mu H_r\right)_{r=\Re^-} = (H_r)_{r=\Re^+} \tag{82}$$

or, based on the first equation (57), as:

$$\left[\mu\left(\frac{\partial \psi}{\partial r}\right) + i\mu_a \frac{1}{\Re}\left(\frac{\partial \psi}{\partial \phi} - \bar{p}\frac{\partial \psi}{\partial \zeta}\right)\right]_{r=\Re^-} = \left(\frac{\partial \psi}{\partial r}\right)_{r=\Re^+}. \tag{83}$$

We are able now to obtain characteristic equations for helical MS waves in a ferrite rod. For the (‡) case we have:

$$(-\mu)^{1/2} \frac{J'_{(w-\bar{p}\beta)}}{J_{(w-\bar{p}\beta)}} + \frac{K'_{(w-\bar{p}\beta)}}{K_{(w-\bar{p}\beta)}} \pm \frac{\mu_a(|w|-\bar{p}|\beta|)}{|\beta|\Re} = 0, \tag{84}$$

where plus corresponds to the situation when both $w$ and $\beta$ are negative and minus – when both $w$ and $\beta$ are positive. For the (±) case we have:

$$(-\mu)^{1/2} \frac{J'_{(w+\bar{p}\beta)}}{J_{(w+\bar{p}\beta)}} + \frac{K'_{(w+\bar{p}\beta)}}{K_{(w+\bar{p}\beta)}} \mp \frac{\mu_a(|w|+\bar{p}|\beta|)}{|\beta|\Re} = 0, \tag{85}$$

where we use minus when $w$ is positive and $\beta$ is negative and plus – when $w$ is negative and $\beta$ is positive. In Eqs. (84) and (85) the prime denotes differentiation with respect to the argument.

The Bessel functions in Eqs. (84) and (85) have orders that are functions of the separation constant $\beta$ along the $\zeta$ direction. The orders are not constant for different values of $\beta$ as it takes place in a cylindrical coordinate system. There is a set of helical waves characterizing by different combinations of signs of $\mu_a$, $w$, $\beta$, and $\bar{p}$. In a smooth infinite ferrite rod, the quantity of pitch $\bar{p}$ is not determined. Since $\bar{p}$ is arbitrary, we have a continuous spectrum of helical MS waves. In particular, one can take formally $\bar{p}=0$. In this case Eqs. (84) and (85) are reduced to the characteristic equation for a ferrite rod in a cylindrical coordinate system (23).

The above model can be characterized as the MS-wave propagation in a helical coordinate system with $\bar{p}>0$. We should also consider the MS-wave propagation in a helical coordinate system with $\bar{p}<0$. It means that together with Eqs. (59) and (60), on has, respectively

$$\frac{\partial^2 \psi}{\partial r^2} + \frac{1}{r}\frac{\partial \psi}{\partial r} + \frac{1}{r^2}\frac{\partial^2 \psi}{\partial \phi^2} + \left(\frac{1}{\mu}+\tan^2\alpha_0\right)\frac{\partial^2 \psi}{\partial \zeta^2} + 2\frac{1}{r}\tan\alpha_0 \frac{\partial^2 \psi}{\partial \phi \partial \zeta} = 0 \tag{59a}$$

and

$$\frac{\partial^2 \psi}{\partial r^2} + \frac{1}{r}\frac{\partial \psi}{\partial r} + \frac{1}{r^2}\frac{\partial^2 \psi}{\partial \phi^2} + \left(1+\tan^2\alpha_0\right)\frac{\partial^2 \psi}{\partial \zeta^2} + 2\frac{1}{r}\tan\alpha_0 \frac{\partial^2 \psi}{\partial \phi \partial \zeta} = 0. \tag{60a}$$

In a helical coordinate system with $\bar{p}<0$ we obtain for the (‡) helical wave

$$\frac{\partial^2 \psi(r)}{\partial r^2} + \frac{1}{r}\frac{\partial \psi(r)}{\partial r} - \left[\frac{\beta^2}{\mu} + \frac{1}{r^2}(w+\bar{p}\beta)^2\right]\psi(r) = 0 \tag{86}$$

inside a ferrite rod $(r \leq \Re)$ and

$$\frac{\partial^2 \psi(r)}{\partial r^2} + \frac{1}{r}\frac{\partial \psi(r)}{\partial r} - \left[\beta^2 + \frac{1}{r^2}(w+\bar{p}\beta)^2\right]\psi(r) = 0 \tag{87}$$

outside a ferrite rod $(r \geq \Re)$. For the (±) helical wave we have

$$\frac{\partial^2 \psi(r)}{\partial r^2} + \frac{1}{r}\frac{\partial \psi(r)}{\partial r} - \left[\frac{\beta^2}{\mu} + \frac{1}{r^2}(w - \overline{p}\beta)^2\right]\psi(r) = 0 \qquad (88)$$

inside a ferrite rod $(r \leq \Re)$ and

$$\frac{\partial^2 \psi(r)}{\partial r^2} + \frac{1}{r}\frac{\partial \psi(r)}{\partial r} - \left[\beta^2 + \frac{1}{r^2}(w - \overline{p}\beta)^2\right]\psi(r) = 0 \qquad (89)$$

outside a ferrite rod $(r \geq \Re)$. Comparing Eqs. (86) – (89) with Eqs. (72), (73), (76), and (77), one can clearly see that the ($\ddagger$) helical wave in a helical coordinate system with $\overline{p} > 0$ corresponds to the ($\pm$) helical wave in a helical coordinate system with $\overline{p} < 0$ and vice versa, the ($\pm$) helical wave in a helical coordinate system with $\overline{p} > 0$ corresponds to the ($\ddagger$) helical wave in a helical coordinate system with $\overline{p} < 0$. In a helical coordinate system with $\overline{p} < 0$, Eqs. (84) and (85) should be rewritten as follows. For the ($\ddagger$) case we have:

$$(-\mu)^{1/2}\frac{J'_{(w+\overline{p}\beta)}}{J_{(w+\overline{p}\beta)}} + \frac{K'_{(w+\overline{p}\beta)}}{K_{(w+\overline{p}\beta)}} \mp \frac{\mu_a(|w| + \overline{p}|\beta|)}{|\beta|\Re} = 0, \qquad (90)$$

where minus corresponds to the situation when both $w$ and $\beta$ are positive and plus – when both $w$ and $\beta$ are negative. For the ($\pm$) case we have:

$$(-\mu)^{1/2}\frac{J'_{(w-\overline{p}\beta)}}{J_{(w-\overline{p}\beta)}} + \frac{K'_{(w-\overline{p}\beta)}}{K_{(w-\overline{p}\beta)}} \pm \frac{\mu_a(|w| - \overline{p}|\beta|)}{|\beta|\Re} = 0, \qquad (91)$$

where we use plus when $w$ is negative and $\beta$ is positive and minus when $w$ is positive and $\beta$ is negative.

For an infinite ferrite rod, in a given coordinate system the ($\ddagger$) helical wave cannot be transformed into the ($\pm$) helical wave and vice versa [25]. The obtained above equations bear a formal character. It is evident that using helical coordinates for the Walker equation (inside a ferrite) and the Laplace equation (outside a ferrite) we have not succeeded to avoid uncertainty in signs of the boundary terms of a ferrite rod. Physically, an analysis of MS waves in an infinite ferrite rod in a helical coordinate system does not have any real sense. This analysis, however, becomes physically sound for a normally magnetized ferrite disk. In a case of a ferrite disk we have two turn points and so solutions become discrete along the $\zeta$ direction. Because of lack of the reflection symmetry on a disk plane (which is underlain by the properties of the Landau-Lifshitz equation) there could be mutual transformations of different helical waves.

## 6. MAGNETOSTATIC HELICAL HARMONICS IN A FERRITE DISK

The above analysis of helical MS waves in a ferrite rod gives evidence that in such coordinates we can *formally* distinguish the waves with clock-wise and counter clock-wise rotation. Also we can distinguish the forward and backward waves. The considered above cases, the ($\ddagger$) case and the ($\pm$) case, provide us with the *basic solutions* for helical waves in the right-handed and left-handed coordinate systems.

To characterize the entire properties of the process we should suppose that MS-potential function is a four-component function:

$$[\psi] = \begin{pmatrix} \psi^{(1)} \\ \psi^{(2)} \\ \psi^{(3)} \\ \psi^{(4)} \end{pmatrix}. \tag{92}$$

Inside a ferrite region ($0 \leq z \leq d, r \leq \Re$) these components are described as:

$$\begin{aligned}
\psi^{(1)} &= a_1 J_{w-\bar{p}\beta}\left[(-\mu)^{1/2} \beta r\right] e^{-iw\phi} e^{-i\beta\zeta}, \\
\psi^{(2)} &= a_2 J_{w+\bar{p}\beta}\left[(-\mu)^{1/2} \beta r\right] e^{+iw\phi} e^{-i\beta\zeta}, \\
\psi^{(3)} &= a_3 J_{w-\bar{p}\beta}\left[(-\mu)^{1/2} \beta r\right] e^{+iw\phi} e^{+i\beta\zeta}, \\
\psi^{(4)} &= a_4 J_{w+\bar{p}\beta}\left[(-\mu)^{1/2} \beta r\right] e^{-iw\phi} e^{+i\beta\zeta}.
\end{aligned} \tag{93}$$

For an outside region $0 \leq z \leq d, r \geq \Re$ one has:

$$\begin{aligned}
\psi^{(1)} &= b_1 K_{w-\bar{p}\beta}(\beta r) e^{-iw\phi} e^{-i\beta\zeta}, \\
\psi^{(2)} &= b_2 K_{w+\bar{p}\beta}(\beta r) e^{+iw\phi} e^{-i\beta\zeta}, \\
\psi^{(3)} &= b_3 K_{w-\bar{p}\beta}(\beta r) e^{+iw\phi} e^{+i\beta\zeta}, \\
\psi^{(4)} &= b_4 K_{w+\bar{p}\beta}(\beta r) e^{-iw\phi} e^{+i\beta\zeta}.
\end{aligned} \tag{94}$$

Coefficients $a_{1,2,3,4}$ and $b_{1,2,3,4}$ in Eqs. (93) and (94) are amplitude coefficients.

Different components of MS-potential function $[\psi]$ are correlated with different wave processes. The main differences between the above four components of function $\psi$ are introduced by factors combining different signs of the quantities $\bar{p}$ and $\mu_a$. Function $\psi^{(1)}$ describes the right-hand-helix MS wave *ascending* in a ferrite disk from $z=0$ to $z=d$ in the helical coordinate system with $\bar{p} > 0$. Function $\psi^{(2)}$ describes the left-hand-helix MS wave *ascending* from $z=0$ to $z=d$ in the helical coordinate system with $\bar{p} > 0$. Function $\psi^{(3)}$ describes the right-hand-helix MS wave *descending* from $z=d$ to $z=0$ in the helical coordinate system with $\bar{p} < 0$. Function $\psi^{(4)}$ is the left-hand-helix MS wave *descending* from $z=d$ to $z=0$ in the helical coordinate system with $\bar{p} < 0$.

Since we have four types of helical MS waves one should consider every space component of a magnetic flux density as a four-component function:

$$[B_r] = \begin{pmatrix} B_r^{(1)} \\ B_r^{(2)} \\ B_r^{(3)} \\ B_r^{(4)} \end{pmatrix}, \quad [B_\phi] = \begin{pmatrix} B_\phi^{(1)} \\ B_\phi^{(2)} \\ B_\phi^{(3)} \\ B_\phi^{(4)} \end{pmatrix}, \quad [B_\zeta] = \begin{pmatrix} B_\zeta^{(1)} \\ B_\zeta^{(2)} \\ B_\zeta^{(3)} \\ B_\zeta^{(4)} \end{pmatrix}. \tag{95}$$

Based on the above characterization the components of the magnetic flux density are expressed as [see Eqs. (57), (84), (85), (90), and (91)]:

$$B_r^{(1)} = -\left[\mu\frac{\partial\psi}{\partial r} + i\mu_a\left(\frac{1}{r}\frac{\partial\psi}{\partial\phi} - \tan\alpha_0\frac{\partial\psi}{\partial\zeta}\right)\right],$$

$$B_\phi^{(1)} = -\frac{1}{\cos\alpha_0}\left[\mu\left(\frac{1}{r}\frac{\partial\psi}{\partial\phi} - \tan\alpha_0\frac{\partial\psi}{\partial\zeta}\right) - i\mu_a\frac{\partial\psi}{\partial r}\right], \quad (96)$$

$$B_\zeta^{(1)} = -\tan\alpha_0\left[\frac{2}{\sin 2\alpha_0}\frac{\partial\psi}{\partial\zeta} - \frac{1}{r}\frac{\partial\psi}{\partial\phi} + (1-\mu)\left(\frac{1}{r}\frac{\partial\psi}{\partial\phi} - \tan\alpha_0\frac{\partial\psi}{\partial\zeta}\right) + i\mu_a\frac{\partial\psi}{\partial r}\right].$$

$$B_r^{(2)} = -\left[\mu\frac{\partial\psi}{\partial r} - i\mu_a\left(\frac{1}{r}\frac{\partial\psi}{\partial\phi} - \tan\alpha_0\frac{\partial\psi}{\partial\zeta}\right)\right],$$

$$B_\phi^{(2)} = -\frac{1}{\cos\alpha_0}\left[\mu\left(\frac{1}{r}\frac{\partial\psi}{\partial\phi} - \tan\alpha_0\frac{\partial\psi}{\partial\zeta}\right) + i\mu_a\frac{\partial\psi}{\partial r}\right], \quad (97)$$

$$B_\zeta^{(2)} = -\tan\alpha_0\left[\frac{2}{\sin 2\alpha_0}\frac{\partial\psi}{\partial\zeta} - \frac{1}{r}\frac{\partial\psi}{\partial\phi} + (1-\mu)\left(\frac{1}{r}\frac{\partial\psi}{\partial\phi} - \tan\alpha_0\frac{\partial\psi}{\partial\zeta}\right) - i\mu_a\frac{\partial\psi}{\partial r}\right].$$

$$B_r^{(3)} = -\left[\mu\frac{\partial\psi}{\partial r} - i\mu_a\left(\frac{1}{r}\frac{\partial\psi}{\partial\phi} + \tan\alpha_0\frac{\partial\psi}{\partial\zeta}\right)\right],$$

$$B_\phi^{(3)} = -\frac{1}{\cos\alpha_0}\left[\mu\left(\frac{1}{r}\frac{\partial\psi}{\partial\phi} + \tan\alpha_0\frac{\partial\psi}{\partial\zeta}\right) + i\mu_a\frac{\partial\psi}{\partial r}\right], \quad (98)$$

$$B_\zeta^{(3)} = \tan\alpha_0\left[-\frac{2}{\sin 2\alpha_0}\frac{\partial\psi}{\partial\zeta} - \frac{1}{r}\frac{\partial\psi}{\partial\phi} + (1-\mu)\left(\frac{1}{r}\frac{\partial\psi}{\partial\phi} + \tan\alpha_0\frac{\partial\psi}{\partial\zeta}\right) - i\mu_a\frac{\partial\psi}{\partial r}\right].$$

$$B_r^{(4)} = -\left[\mu\frac{\partial\psi}{\partial r} + i\mu_a\left(\frac{1}{r}\frac{\partial\psi}{\partial\phi} + \tan\alpha_0\frac{\partial\psi}{\partial\zeta}\right)\right],$$

$$B_\phi^{(4)} = -\frac{1}{\cos\alpha_0}\left[\mu\left(\frac{1}{r}\frac{\partial\psi}{\partial\phi} + \tan\alpha_0\frac{\partial\psi}{\partial\zeta}\right) - i\mu_a\frac{\partial\psi}{\partial r}\right], \quad (99)$$

$$B_\zeta^{(4)} = \tan\alpha_0\left[-\frac{2}{\sin 2\alpha_0}\frac{\partial\psi}{\partial\zeta} - \frac{1}{r}\frac{\partial\psi}{\partial\phi} + (1-\mu)\left(\frac{1}{r}\frac{\partial\psi}{\partial\phi} + \tan\alpha_0\frac{\partial\psi}{\partial\zeta}\right) + i\mu_a\frac{\partial\psi}{\partial r}\right].$$

Eqs. (96) – (99) show that for different combinations of signs of $\bar{p}$ and $\mu_a$ one has different fields of the magnetic flux density. There are four types of fields.

We can see for ourselves now that components $\psi^{(1)}$ and $\psi^{(4)}$ are described by the characteristic equation of one type. On the other hand, components $\psi^{(2)}$ and $\psi^{(3)}$ are also described by the characteristic equation of one type (which is not the same as a characteristic equation for components $\psi^{(1)}$ and $\psi^{(4)}$). Really, for functions $\psi^{(1)}$ and $\psi^{(4)}$ the characteristic equation is [see Eq. (84) and (91) with the corresponding signs]:

$$(-\mu)^{1/2}\frac{J'_{(w-\bar{p}\beta)}}{J_{(w-\bar{p}\beta)}} + \frac{K'_{(w-\bar{p}\beta)}}{K_{(w-\bar{p}\beta)}} + \frac{\mu_a(|w|-\bar{p}|\beta|)}{|\beta|\mathfrak{R}} = 0. \quad (100)$$

For functions $\psi^{(2)}$ and $\psi^{(3)}$ we have [see Eq. (85) and (90) with the corresponding signs]:

$$(-\mu)^{1/2}\frac{J'_{(w+\bar{p}\beta)}}{J_{(w+\bar{p}\beta)}}+\frac{K'_{(w+\bar{p}\beta)}}{K_{(w+\bar{p}\beta)}}-\frac{\mu_a(|w|+\bar{p}|\beta|)}{|\beta|\Re}=0. \tag{101}$$

From the above characteristic equations it becomes clear that we have the same moduli of the propagation constants for helical modes $\psi^{(1)}$ and $\psi^{(4)}$ as well as the same moduli of the propagation constants for helical modes $\psi^{(2)}$ and $\psi^{(3)}$. At the same time, propagation constants of modes $\psi^{(1)}$ and $\psi^{(4)}$ are different from the propagation constants of modes $\psi^{(2)}$ and $\psi^{(3)}$. So we can write:

$$\left|w^{(1)}\right|=\left|w^{(4)}\right|\neq\left|w^{(2)}\right|=\left|w^{(3)}\right| \tag{102}$$

and

$$\left|\beta^{(1)}\right|=\left|\beta^{(4)}\right|\neq\left|\beta^{(2)}\right|=\left|\beta^{(3)}\right|. \tag{103}$$

Taking these relations into account, one can see that for functions $\psi^{(1)}$ and $\psi^{(4)}$ there are the same Bassel equations inside and outside a ferrite [see, respectively, Eqs. (72) and (73)]:

$$\frac{\partial^2\psi^{(1,4)}}{\partial r^2}+\frac{1}{r}\frac{\partial\psi^{(1,4)}}{\partial r}-\left[\frac{\left(\beta^{(1,4)}\right)^2}{\mu}+\frac{1}{r^2}\left(w^{(1,4)}-\bar{p}\beta^{(1,4)}\right)^2\right]\psi^{(1,4)}=0 \tag{104}$$

and

$$\frac{\partial^2\psi^{(1,4)}}{\partial r^2}+\frac{1}{r}\frac{\partial\psi^{(1,4)}}{\partial r}-\left[\left(\beta^{(1,4)}\right)^2+\frac{1}{r^2}\left(w^{(1,4)}-\bar{p}\beta^{(1,4)}\right)^2\right]\psi^{(1,4)}=0. \tag{105}$$

Similarly, for functions $\psi^{(2)}$ and $\psi^{(3)}$ we have [see Eqs. (76) and (77)]:

$$\frac{\partial^2\psi^{(2,3)}}{\partial r^2}+\frac{1}{r}\frac{\partial\psi^{(2,3)}}{\partial r}-\left[\frac{\left(\beta^{(2,3)}\right)^2}{\mu}+\frac{1}{r^2}\left(w^{(2,3)}+\bar{p}\beta^{(2,3)}\right)^2\right]\psi^{(2,3)}=0 \tag{106}$$

and

$$\frac{\partial^2\psi^{(2,3)}}{\partial r^2}+\frac{1}{r}\frac{\partial\psi^{(2,3)}}{\partial r}-\left[\left(\beta^{(2,3)}\right)^2+\frac{1}{r^2}\left(w^{(2,3)}+\bar{p}\beta^{(2,3)}\right)^2\right]\psi^{(2,3)}=0. \tag{107}$$

On planes $z = 0, d$ reflections of helical MS wave propagating inside a ferrite disk take place. For these helical waves one has the failure of the law of reflection symmetry. It means that any helical wave "incident" on a reflection plane should be transformed to another-type helical wave. In other words, on a reflection plane one has coupling between different helical waves (between waves with different types of symmetry). Suppose, for example, that we have a helical MS mode $\psi^{(1)}$ incident (from a ferrite region) on a plane $z = d$. On the plane $z = d$ this ascending helical wave can be coupled with descending helical MS modes $\psi^{(3)}$ and (or) $\psi^{(4)}$. At the same time a helical MS mode $\psi^{(2)}$ incident on a plane $z = d$ can be coupled with helical MS modes $\psi^{(3)}$ and (or) $\psi^{(4)}$. Similarly, due to reflections on a plane $z=0$ descending mode $\psi^{(3)}$ becomes coupled with ascending modes $\psi^{(1)}$ and (or) $\psi^{(2)}$. Also mode $\psi^{(4)}$ becomes coupled with modes $\psi^{(1)}$ and (or) $\psi^{(2)}$. So in a general consideration there should be a system of four linear equations for four variables: $\psi^{(1)}, \psi^{(2)}, \psi^{(3)}$, and $\psi^{(4)}$. The above analysis shows, however, that there could be two special cases of the *resonance interactions*. The first resonant case one has due to ascending helical wave $\psi^{(1)}$ and descending helical wave $\psi^{(4)}$. The second resonant case is due to ascending helical wave $\psi^{(2)}$ and descending helical wave $\psi^{(3)}$. These resonant cases, which we will denote conventionally as the "right" and "left" resonances, are shown, respectively, in Figs. 1a and 1b.

In spite of the reflection breaking symmetry in turn points of a flat ferrite disk, it is evident, however, that for different-type helical harmonic there should be certain symmetry properties with respect to a disk radius. Among different kinds of possible symmetry relations the most interesting and important are the following ones:

$$\frac{\partial J^{(1)}}{\partial r} = \frac{\partial J^{(4)}}{\partial r} \tag{108}$$

and

$$\frac{\partial J^{(2)}}{\partial r} = \frac{\partial J^{(3)}}{\partial r}, \tag{109}$$

where

$$J^{(1)} \equiv J_{w^{(1)} - \bar{p}\beta^{(1)}}\left[(-\mu)^{1/2}\beta^{(1)} r\right], \quad J^{(4)} \equiv J_{w^{(4)} + \bar{p}\beta^{(4)}}\left[(-\mu)^{1/2}\beta^{(4)} r\right] \tag{110}$$

and

$$J^{(2)} \equiv J_{w^{(2)} + \bar{p}\beta^{(2)}}\left[(-\mu)^{1/2}\beta^{(2)} r\right], \quad J^{(3)} \equiv J_{w^{(3)} - \bar{p}\beta^{(3)}}\left[(-\mu)^{1/2}\beta^{(3)} r\right]. \tag{111}$$

The relations are written for any coordinate $r$ $(0 \leq r \leq \Re)$.

The known recurrence formulae for the Bessel functions allow us to rewrite Eqs. (108) and (109) in a following form:

$$\frac{w - \bar{p}\beta}{x} J_{w-\bar{p}\beta}(x) - J_{w-\bar{p}\beta+1}(x) = \frac{w + \bar{p}\beta}{x} J_{w+\bar{p}\beta}(x) - J_{w+\bar{p}\beta+1}(x), \tag{112}$$

where $x \equiv (-\mu)^{1/2} \beta r$. It is more relevant to rewrite Eq. (112) in the integral form:

$$\int_{x=0}^{x=(-\mu)^{1/2}\beta\Re} \left[\frac{w - \bar{p}\beta}{x} J_{w-\bar{p}\beta}(x) - J_{w-\bar{p}\beta+1}(x)\right] dx = \int_{x=0}^{x=(-\mu)^{1/2}\beta\Re} \left[\frac{w + \bar{p}\beta}{x} J_{w+\bar{p}\beta}(x) - J_{w+\bar{p}\beta+1}(x)\right] dx. \tag{113}$$

Here we omitted for simplicity the corresponding superscripts of numbers of $w$ and $\beta$.

For a given quantity of pitch $\bar{p}$, boundary integral equation (113) together with boundary differential equations (100) and (101) are the systems of equations which allow finding parameters of helical harmonics. The quantities of pitch $\bar{p}$ one has from the evident resonance conditions:

$$p = \frac{d}{2}n, \qquad n = 1, 2, 3, \ldots \tag{114}$$

So

$$\tan \alpha_0 = \frac{p}{2\pi\Re} = \frac{d}{4\pi\Re} n. \tag{115}$$

One of the interesting conclusion, we can made based on symmetry relations (108) and (109), arises from the following consideration. To have mutual transformations of helical modes in the reflection points, stipulating the above two types of resonances ($\psi^{(1)} \leftrightarrow \psi^{(4)}$ and $\psi^{(2)} \leftrightarrow \psi^{(3)}$), one should suppose also that in the turn points on the reflection surfaces there are continuous quantities of the corresponding components $B_r$ and $B_\theta$ of the magnetic flux density. It means that in the reflection points there should be

$$B_r^{(1)} = B_r^{(4)}, \quad B_\theta^{(1)} = B_\theta^{(4)} \tag{116}$$

and

$$B_r^{(2)} = B_r^{(3)}, \quad B_\theta^{(2)} = B_\theta^{(3)}. \tag{117}$$

Taking into account Eqs. (96), (99), (102), and (103), and relations between cylindrical and helical components [23], $B_\theta = B_\phi \cos\alpha_0$, one can see that equalities (116) are valid if the symmetry relation (108) takes place. Similarly, with use of Eqs. (97), (98), (102), and (103) it becomes evident that equalities (117) will be valid if the symmetry relation (109) occurs. Another interesting conclusion regarding relations (108) and (109) we will make below when analyzing the power flows in a ferrite disk.

In regions above and below a ferrite disk ($z \leq 0, z \geq d$) there are exponentially descending solutions along $z$ axis. Only with such an evanescent assumption we will be able to get an analytical solution of the problem. We do not preserve the helical-type properties of MS functions outside a ferrite disk. So in the outside regions we should put $\bar{p} = 0$. In the outside regions $z \leq 0, z \geq d$ for any $z=const$ we have the similar-type "flat patterns" for MS-potential function distribution. Following the method of separation of variables used in [11], in regions $z \leq 0, z \geq d$ and for $r \leq \Re$ we describe the MS-potential function by the Bessel equation:

$$\frac{\partial^2 \psi(r)}{\partial r^2} + \frac{1}{r}\frac{\partial \psi(r)}{\partial r} + \left[\alpha^2 - \frac{\nu^2}{r^2}\right]\psi(r) = 0. \tag{118}$$

MS-potential function in the outside region $z \geq d, r \leq \Re$ is described as:

$$\psi = f_1 e^{-i\nu\theta} e^{-\alpha(z-d)} \tag{119}$$

and in the outside region $z \leq 0, r \leq \Re$ is described as:

$$\psi = f_2 e^{-i\nu\theta} e^{\alpha z}. \tag{120}$$

Coefficients $f_1$ and $f_2$ are amplitude coefficients.

Symmetry breaking on plane surfaces of a ferrite disk leading to mutual transformations of different-type helical modes ($\psi^{(1)} \leftrightarrow \psi^{(4)}$ and $\psi^{(2)} \leftrightarrow \psi^{(3)}$) should not cause discontinuities of MS-potential functions and normal components of the magnetic flux density at the reflection points. The boundary conditions at the reflection points on plane surfaces of a ferrite disk are:

$$\psi^{(1)}_{\substack{z=0^+ \\ z=d^-}} + \psi^{(4)}_{\substack{z=0^+ \\ z=d^-}} = \psi^{(R)}_{\substack{z=0^- \\ z=d^+}}, \qquad \left(B_z\right)^{(1)}_{\substack{z=0^+ \\ z=d^-}} + \left(B_z\right)^{(4)}_{\substack{z=0^+ \\ z=d^-}} = \left(B_z\right)^{(R)}_{\substack{z=0^- \\ z=d^+}}, \qquad (121)$$

and

$$\psi^{(2)}_{\substack{z=0^+ \\ z=d^-}} + \psi^{(3)}_{\substack{z=0^+ \\ z=d^-}} = \psi^{(L)}_{\substack{z=0^- \\ z=d^+}}, \qquad \left(B_z\right)^{(2)}_{\substack{z=0^+ \\ z=d^-}} + \left(B_z\right)^{(3)}_{\substack{z=0^+ \\ z=d^-}} = \left(B_z\right)^{(L)}_{\substack{z=0^- \\ z=d^+}}, \qquad (122)$$

where superscripts (*R*) and (*L*) correspond to the "right" and "left" resonances.

We define now a four-component-function of a space *z*-component of a magnetic flux density. Since in a ferrite region $B_z = B_\zeta + B_\phi \sin\alpha_0$ [23], one has from Eqs. (96) – (99) after some algebraic transformations:

$$B_z^{(1,2,3,4)} = -\frac{\partial \psi^{(1,2,3,4)}}{\partial \zeta} = -\frac{\partial \psi^{(1,2,3,4)}}{\partial z}. \qquad (123)$$

One can rewrite this equation as:

$$[B_z] = -\begin{pmatrix} \nabla_\| \psi^{(1)} \\ \nabla_\| \psi^{(2)} \\ \nabla_\| \psi^{(3)} \\ \nabla_\| \psi^{(4)} \end{pmatrix}. \qquad (124)$$

The boundary conditions (121) and (122) can be rewritten as:

$$\psi^{(1)}_{\substack{z=0^+ \\ z=d^-}} + \psi^{(4)}_{\substack{z=0^+ \\ z=d^-}} = \psi^{(R)}_{\substack{z=0^- \\ z=d^+}}, \qquad \left(\frac{\partial \psi^{(1)}}{\partial z}\right)_{\substack{z=0^+ \\ z=d^-}} + \left(\frac{\partial \psi^{(4)}}{\partial z}\right)_{\substack{z=0^+ \\ z=d^-}} = \left(\frac{\partial \psi^{(R)}}{\partial z}\right)_{\substack{z=0^- \\ z=d^+}}, \qquad (125)$$

and

$$\psi^{(2)}_{\substack{z=0^+ \\ z=d^-}} + \psi^{(3)}_{\substack{z=0^+ \\ z=d^-}} = \psi^{(L)}_{\substack{z=0^- \\ z=d^+}}, \qquad \left(\frac{\partial \psi^{(2)}}{\partial z}\right)_{\substack{z=0^+ \\ z=d^-}} + \left(\frac{\partial \psi^{(3)}}{\partial z}\right)_{\substack{z=0^+ \\ z=d^-}} = \left(\frac{\partial \psi^{(L)}}{\partial z}\right)_{\substack{z=0^- \\ z=d^+}}. \qquad (126)$$

## 7. POWER FLOW RELATIONS FOR MAGNETOSTATIC HELICAL HARMONICS IN A FERRITE DISK

For MS-wave processes a power flow density is expressed as [11]:

$$\vec{P} = \frac{i\omega}{4}\left(\psi \vec{B}^* - \psi^* \vec{B}\right). \qquad (127)$$

To consider a power flow density we should take into account the above four-component functions. For mode $\psi^{(1)}$ we have from Eqs. (93) and (96):

$$P_\phi^{(1)} = \frac{i\omega}{4}\left[\psi^{(1)}\left(B_\phi^{(1)}\right)^* - \left(\psi^{(1)}\right)^* B_\phi^{(1)}\right] = A_\phi^{(1)}\left[\mu\left(\frac{1}{r}w^{(1)} - \beta^{(1)}\tan\alpha_0\right) + \mu_a \frac{\partial J^{(1)}}{\partial r}\right], \quad (128)$$

$$P_\zeta^{(1)} = \frac{i\omega}{4}\left[\psi^{(1)}\left(B_\zeta^{(1)}\right)^* - \left(\psi^{(1)}\right)^* B_\zeta^{(1)}\right] = \\ A_\zeta^{(1)}\left[\frac{2}{\sin 2\alpha_0}\beta^{(1)} - \frac{1}{r}w^{(1)} + (1-\mu)\left(\frac{1}{r}w^{(1)} - \beta^{(1)}\tan\alpha_0\right) - \mu_a \frac{\partial J^{(1)}}{\partial r}\right], \quad (129)$$

where $A_\phi^{(1)}$ and $A_\zeta^{(1)}$ are real coefficients. Similarly, based on Eqs. (93) and (99), one obtains for mode $\psi^{(4)}$:

$$P_\phi^{(4)} = \frac{i\omega}{4}\left[\psi^{(4)}\left(B_\phi^{(4)}\right)^* - \left(\psi^{(4)}\right)^* B_\phi^{(4)}\right] = A_\phi^{(4)}\left[\mu\left(\frac{1}{r}w^{(4)} - \beta^{(4)}\tan\alpha_0\right) + \mu_a \frac{\partial J^{(4)}}{\partial r}\right], \quad (130)$$

$$P_\zeta^{(4)} = \frac{i\omega}{4}\left[\psi^{(4)}\left(B_\zeta^{(4)}\right)^* - \left(\psi^{(4)}\right)^* B_\zeta^{(4)}\right] = \\ A_\zeta^{(4)}\left[\frac{2}{\sin 2\alpha_0}\beta^{(4)} - \frac{1}{r}w^{(4)} + (1-\mu)\left(\frac{1}{r}w^{(4)} - \beta^{(4)}\tan\alpha_0\right) - \mu_a \frac{\partial J^{(4)}}{\partial r}\right], \quad (131)$$

where $A_\phi^{(4)} = A_\phi^{(1)}$ and $A_\zeta^{(4)} = -A_\zeta^{(1)}$.

One can see that if relations (108) and (109) are valid, the following equalities take place:

$$P_\phi^{(1)} = P_\phi^{(4)} \quad \text{and} \quad P_\zeta^{(1)} = -P_\zeta^{(4)} \quad (132)$$

By a similar way we find proper relations for quantities $P_\phi^{(2)}, P_\zeta^{(2)}, P_\phi^{(3)}$, and $P_\zeta^{(3)}$. One can see that the equalities

$$P_\phi^{(2)} = P_\phi^{(3)} \quad \text{and} \quad P_\zeta^{(2)} = -P_\zeta^{(3)} \quad (133)$$

will be valid if symmetry relations (108) and (109) are relevant.

In the Waldron's helical coordinate system [23], line elements are expressed as:

$$ds_1 = dr, \quad ds_2 = \sqrt{r^2 + \bar{p}^2}\,d\phi, \quad ds_3 = d\zeta$$

and a distance element as:

$$ds = \sqrt{dr^2 + (r^2 + \bar{p}^2)\,d\phi^2 + d\zeta^2 + 2\bar{p}\,d\phi\,d\zeta}. \quad (134)$$

Based on the last equation we can obtain an expression for a power flow. In a helical coordinate system the MS-wave power flow propagates along the $\phi$- and $\zeta$- coordinates. For any helical mode $n$ ($n = 1, 2, 3, 4$) with the corresponding power flow coordinates $P_\phi^{(n)}$ and $P_\zeta^{(n)}$, the total quantity of the power flow is expressed as:

$$P^{(n)} = \sqrt{(r^2 + \bar{p}^2)\left(P_\phi^{(n)}\right)^2 + \left(P_\zeta^{(n)}\right)^2 + 2\bar{p}^2 P_\phi^{(n)} P_\zeta^{(n)}} \ . \tag{135}$$

One can see from Eqs. (132) and (135) that for the "right" resonance (the $\psi^{(1)} \leftrightarrow \psi^{(4)}$ interaction) there is an inequality of the total eigen power flows:

$$\left|P^{(1)}\right| \neq \left|P^{(4)}\right|. \tag{136}$$

Similarly, Eqs. (133) and (135) give an inequality of the total eigen power flows for the "left" resonance (the $\psi^{(2)} \leftrightarrow \psi^{(3)}$ interaction):

$$\left|P^{(2)}\right| \neq \left|P^{(3)}\right|. \tag{137}$$

Such inequalities for total eigen power flows is an enough evident fact, since the helical coordinate system is a non orthogonal system of coordinates. Two resonances (the "right" and "left" resonances) are not orthogonal with respect to energy.

To get a complete picture of power flows of magnetostatic helical modes in a ferrite disk, one should consider the four-component vectors (92) and (95). It means that in a general case together with the interactions $\psi^{(1)} \leftrightarrow \psi^{(4)}$ and $\psi^{(2)} \leftrightarrow \psi^{(3)}$ we have to take into account the interactions $\psi^{(1)} \leftrightarrow \psi^{(3)}$ and $\psi^{(2)} \leftrightarrow \psi^{(4)}$. Contrary to the continuity relations (116) and (117), in the last case one has the following inequalities in the reflection points:

$$B_r^{(1)} \neq B_r^{(3)}, \ B_\theta^{(1)} \neq B_\theta^{(3)} \tag{138}$$

and

$$B_r^{(2)} \neq B_r^{(4)}, \ B_\theta^{(2)} \neq B_\theta^{(4)} \ . \tag{139}$$

The differences between radial and azimuth components of the magnetic flux density are expressed as:

$$B_r^{(3)} - B_r^{(1)} = \mu\left(\frac{\partial \psi^{(3)}}{\partial r} - \frac{\partial \psi^{(1)}}{\partial r}\right) - i\mu_a \tan\alpha_0 \left(\frac{\partial \psi^{(3)}}{\partial \zeta} - \frac{\partial \psi^{(1)}}{\partial \zeta}\right) - i\mu_a \frac{1}{r}\left(\frac{\partial \psi^{(3)}}{\partial \phi} + \frac{\partial \psi^{(1)}}{\partial \phi}\right) \tag{140}$$

and

$$B_\theta^{(3)} - B_\theta^{(1)} = \mu\frac{1}{r}\left(\frac{\partial \psi^{(3)}}{\partial \phi} - \frac{\partial \psi^{(1)}}{\partial \phi}\right) + \mu \tan\alpha_0 \left(\frac{\partial \psi^{(3)}}{\partial \zeta} + \frac{\partial \psi^{(1)}}{\partial \zeta}\right) + i\mu_a \left(\frac{\partial \psi^{(3)}}{\partial r} + \frac{\partial \psi^{(1)}}{\partial r}\right). \tag{141}$$

The right-hand side of Eq. (140) is a real quantity and the right-hand side of Eq. (141) is an imaginary quantity.

## 8. THE DIRAC-LIKE EQUATION FOR MAGNETIC DIPOLAR MODES

Since helical coordinates are not separable, to get the physically adequate models for MS oscillations in a ferrite disk we have to correlate the obtained results with the ones given from the cylindrical coordinate system. If in a structure under consideration one does not distinguish the left- or right-handedness, the results obtained in helical coordinates will be the same as in cylindrical coordinates. When a structure demonstrates the handedness properties, the physical models in cylindrical and helical coordinates will be different. The handedness of magnetic-dipolar modes in a ferrite disk becomes evident in a helical coordinate system because of certain correlations between signs of pitch $\bar{p}$ and the component $\mu_a$.

In the above analysis we showed that in a helical coordinate system we recognize the four-component MS-potential wave functions $[\psi]$. If now we come back to a cylindrical coordinate system we should also distinguish four scalar-wave solutions which we denote as $[\breve{\psi}]$. In a general consideration, the four-component scalar-wave process in a cylindrical coordinate system should be expressed as a system of four first-order differential equations:

$$[I]\frac{\partial [\breve{\psi}]}{\partial t} = [A]\frac{\partial [\breve{\psi}]}{\partial r} + [B]\frac{\partial [\breve{\psi}]}{\partial \theta} + [C]\frac{\partial [\breve{\psi}]}{\partial z} + [D]\breve{\psi}, \qquad (142)$$

where $[I]$ is the four-order unit matrix, $[D]$ is the four-order diagonal matrix and $[A]$, $[B]$, $[C]$ are four-order matrices.

A dipole-dipole interaction between the magnetic moments of the atoms is considered as purely relativistic in origin. At the same time, magnetic-dipolar oscillating modes are not classical electromagnetic waves. In numerous publications of the MS-wave theory (see e.g. [6,7] and references therein]), it has been stated that magnetic-dipolar modes are only the approximation solutions of "pure" electromagnetic-wave oscillations. As we discuss in this section, the MS-wave description is really an approximation related, however, not to the Maxwell equations but to the Dirac-like equation. The question arises: Is it possible to form the Dirac-like relativistic-invariant expression for magnetic-dipolar oscillating modes?

If one supposes formally that in the above expressions $\bar{p}=0$ and $\mu_a=0$, the helical states become degenerate: $\psi^{(1)} = \psi^{(2)} = \psi^{(3)} = \psi^{(4)} \equiv \psi$. In this case the function $\psi$ is described by the Schrödinger-like equation and one has the complete-set *energy-eigenstate* spectrum for magnetic-dipolar modes [11,12]. When $\bar{p} \neq 0$ and $\mu_a \neq 0$, in a helical coordinate system the Schrödinger-like equation written for one-component function $\psi$ becomes "split" to the equation written for four-component function $[\psi]$. It leads to the fact that in a cylindrical coordinate system instead of helical harmonics we obtain spinor wave functions.

The "right" and "left" resonances, corresponding to the interactions $\psi^{(1)} \leftrightarrow \psi^{(4)}$ and $\psi^{(2)} \leftrightarrow \psi^{(3)}$ (see Figs. 1 a, b), we have when integer numbers *n* in Eq. (114) are *even* quantities. In this case a total period of $2\pi$ rotation in a cylindrical coordinate system corresponds (for the same time) to the $4\pi$ rotation in a helical coordinate system. This is the model of spinning rotation described (based on another type of an analysis) in Section 4 of the paper. Considering the interactions $\psi^{(1)} \leftrightarrow \psi^{(4)}$ in a cylindrical coordinates one sees that time variation of function $\breve{\psi}^{(1)}$ is due to interaction with function $\breve{\psi}^{(4)}$ via mutual spinning rotation. Similarly, time variation of function $\breve{\psi}^{(2)}$ is due to interaction with function $\breve{\psi}^{(3)}$ via mutual spinning rotation. For these interactions we can rewrite Eq. (142) as two systems of two equations:

$$iX\frac{\partial \breve{\psi}^{(1)}}{\partial t} = Ye^{-i\theta}\left(\frac{\partial}{\partial r} - i\frac{1}{r}\frac{\partial}{\partial \theta}\right)\breve{\psi}^{(4)},$$

$$iX\frac{\partial \breve{\psi}^{(4)}}{\partial t} = Ye^{+i\theta}\left(\frac{\partial}{\partial r} + i\frac{1}{r}\frac{\partial}{\partial \theta}\right)\breve{\psi}^{(1)}$$

(143)

and

$$iX\frac{\partial \breve{\psi}^{(2)}}{\partial t} = Ye^{+i\theta}\left(\frac{\partial}{\partial r} + i\frac{1}{r}\frac{\partial}{\partial \theta}\right)\breve{\psi}^{(3)},$$

$$iX\frac{\partial \breve{\psi}^{(3)}}{\partial t} = Ye^{-i\theta}\left(\frac{\partial}{\partial r} - i\frac{1}{r}\frac{\partial}{\partial \theta}\right)\breve{\psi}^{(2)},$$

(144)

where $X$ and $Y$ are constant coefficients. These equations can be transformed in the Cartesian coordinate system with use of the notation

$$\frac{\partial}{\partial x} \pm i\frac{\partial}{\partial y} = e^{\pm i\theta}\left(\frac{\partial}{\partial r} \pm i\frac{1}{r}\frac{\partial}{\partial \theta}\right).$$

(145)

Eqs. (143) and (144) are the Dirac-like equations for massless fermions written in a cylindrical coordinate system. Operator $iX\frac{\partial}{\partial t}$ is the energy operator $E$ (like it was for function $\psi$ described by the Schrödinger-like equation [11, 12]). Coefficient $X$ has the meaning of the "effective Plank constant" and coefficient $Y$ is the "effective speed of light". The values $X$ and $Y$ can be found from the dispersion characteristics of MS modes. For a monochromatic process Eqs. (143) and (144) describes a certain state. For every given state there are certain coefficients $X$ and $Y$.

The importance of the above identification with the Dirac equation is that it immediately permits to construct the azimuthal parts of the spinor wave functions. Based on coordinate transformation between helical and cylindrical systems [23], one can see that in a cylindrical coordinate system, Eqs. (100) and (101) take a form for the "right" and "left" resonances, respectively, as

$$(-\mu)^{\frac{1}{2}}\left(\frac{J'_{\nu_R}}{J_{\nu_R}}\right)_{r=\Re} + \left(\frac{K'_{\nu_R}}{K_{\nu_R}}\right)_{r=\Re} + \frac{\mu_a \nu_R}{|\beta_R|\Re} = 0$$

(146)

and

$$(-\mu)^{\frac{1}{2}}\left(\frac{J'_{\nu_L}}{J_{\nu_L}}\right)_{r=\Re} + \left(\frac{K'_{\nu_L}}{K_{\nu_L}}\right)_{r=\Re} - \frac{\mu_a \nu_L}{|\beta_L|\Re} = 0.$$

(147)

These equations have a form of Eq. (23). Quantities $\nu_R$ and $\nu_L$ are integer numbers. The eigenvalues of the "total angular moment" in the Dirac coordinate system are half-integer and doubly degenerate with the eigenspinors of $j = \nu + \frac{1}{2}$ given by

$$\begin{pmatrix} e^{i\nu\theta} \\ 0 \end{pmatrix}$$

and

$$\begin{pmatrix} 0 \\ e^{i(\nu+1)\theta} \end{pmatrix}$$

These states have opposite angular momenta. In this case an analysis can be reduced to consideration of a system of coupled ordinary differential equations [26].

Helical modes in a ferrite disk are accompanied with surface magnetic currents. There are currents caused by boundary conditions on a lateral surface of a ferrite disk. A type of a current is correlated with a type of the helical mode interaction. For interactions $\psi^{(1)} \leftrightarrow \psi^{(4)}$, $\psi^{(2)} \leftrightarrow \psi^{(3)}$ surface magnetic currents represent a continuous double helix. In this case an anapole moment takes place. The symmetry properties taking into account the anapole moment orientations are shown in paper [22]. Such properties may stipulate interaction with the external normal RF electric field [40].

The "right" and "left" resonances become coupled for a ferrite disk placed in a homogeneous tangential RF magnetic field [15, 16]. One also observes such resonance coupling for a ferrite disk with a symmetrically oriented linear surface electrode, when this ferrite particle is placed in a homogeneous tangential RF electric field. Both these cases are experimentally observed [17]. There are different ways of combining the right and left moving waves in order to create standing waves. One of the possibilities corresponds to situation when time variation of function $\breve{\psi}^{(1)}$ is due to interaction with function $\breve{\psi}^{(4)}$ via mutual spinning rotation and with function $\breve{\psi}^{(3)}$ by mutual linear motion. Also time variation of function $\breve{\psi}^{(2)}$ is due to interaction with function $\breve{\psi}^{(3)}$ via mutual spinning rotation and with function $\breve{\psi}^{(4)}$ by mutual linear motion. For such mix interaction we have

$$\begin{aligned} iX\frac{\partial \breve{\psi}^{(1)}}{\partial t} &= Ye^{-i\theta}\left(\frac{\partial}{\partial r} - i\frac{1}{r}\frac{\partial}{\partial \theta}\right)\breve{\psi}^{(4)} + Y\frac{\partial \breve{\psi}^{(3)}}{\partial z} + Z\breve{\psi}^{(1)}, \\ iX\frac{\partial \breve{\psi}^{(2)}}{\partial t} &= Ye^{+i\theta}\left(\frac{\partial}{\partial r} + i\frac{1}{r}\frac{\partial}{\partial \theta}\right)\breve{\psi}^{(3)} - Y\frac{\partial \breve{\psi}^{(4)}}{\partial z} + Z\breve{\psi}^{(2)}, \\ iX\frac{\partial \breve{\psi}^{(3)}}{\partial t} &= Ye^{-i\theta}\left(\frac{\partial}{\partial r} - i\frac{1}{r}\frac{\partial}{\partial \theta}\right)\breve{\psi}^{(2)} + Y\frac{\partial \breve{\psi}^{(1)}}{\partial z} - Z\breve{\psi}^{(3)}, \\ iX\frac{\partial \breve{\psi}^{(4)}}{\partial t} &= Ye^{+i\theta}\left(\frac{\partial}{\partial r} + i\frac{1}{r}\frac{\partial}{\partial \theta}\right)\breve{\psi}^{(1)} - Y\frac{\partial \breve{\psi}^{(2)}}{\partial z} - Z\breve{\psi}^{(4)}. \end{aligned} \quad (148)$$

Coefficient Z in Eqs. (148) is the "Dirac effective mass".

A detail analysis of the Dirac-like equation for magnetic-dipolar modes and numerical evaluations are beyond the scope of this article and is the subject of our future publications.

## 9. CONCLUSION

The spectral properties of magnetic-dipolar modes in a ferrite disk resonator show that these modes are neither electromagnetic nor exchange-interaction waves. As a distinctive feature of these modes there is the reflection symmetry breaking which leads to appearance of four types of helical harmonics for magnetostatic-potential wave functions. This takes place due to the presence of surface magnetic currents which are chiral currents.

We analyzed the spinor wave functions of magnetic-dipolar modes in a ferrite disk by two ways. Initially, in Section 4, we showed that the properties of a boundary magnetic current are described based on double-valued magnetostatic-potential functions. This provides us with such physical notions as "electric spins" (pseudospins) and anapole moments. The "spinning coordinates" used in such a description are singular coordinates on a ferrite-disk lateral surface. Another consideration, based on an analysis of helical harmonics, leads us to the Dirac-like equations with further analysis of the azimuthal parts of the spinor wave functions.

In a normally magnetized ferrite disk the "electric spin" (pseudospin) of a magnetic-dipolar mode is tied to the anapole linear momentum $\vec{a}^e$. This is completely analogous to the physical spin of a massless neutrino which points along the direction of propagation. So in our case one can distinguish the "particles" (when the pseudospin is antiparallel to the anapole moment) and "aniparticles" (when the pseudospin is parallel to the anapole moment).

**Figure captions**

Fig. 1 The "right" and "left" resonances.

(a) The "right" resonance caused by the $\psi^{(1)} \leftrightarrow \psi^{(4)}$ interaction (Arrows show directions of propagation for helical MS modes).

(b) The "left" resonance caused by the $\psi^{(2)} \leftrightarrow \psi^{(3)}$ interaction (Arrows show directions of propagation for helical MS modes).

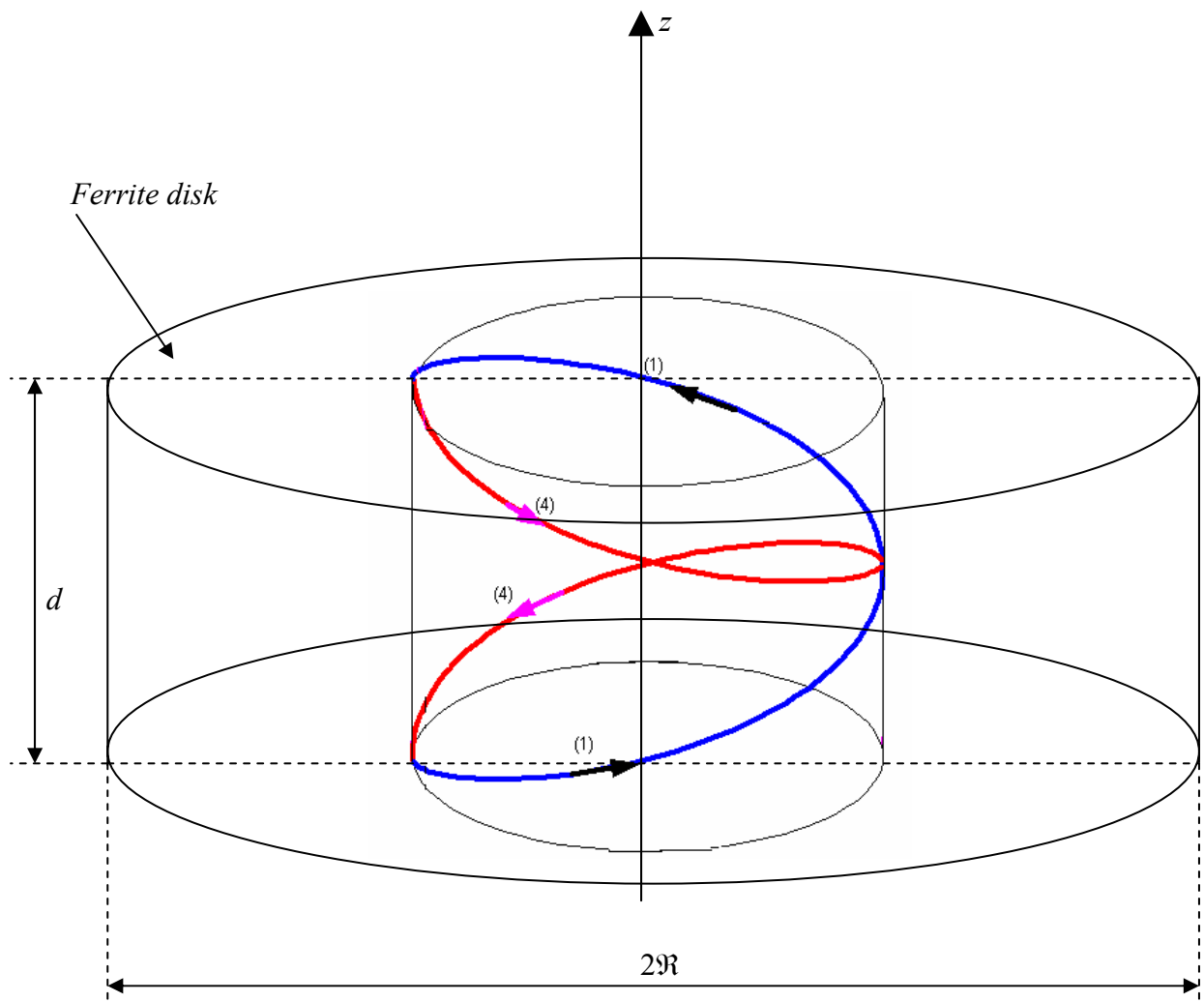

Fig 1a: The "right" resonance caused by the $\psi^{(1)} \leftrightarrow \psi^{(4)}$ interaction (arrows show directions of propagation for helical MS modes)

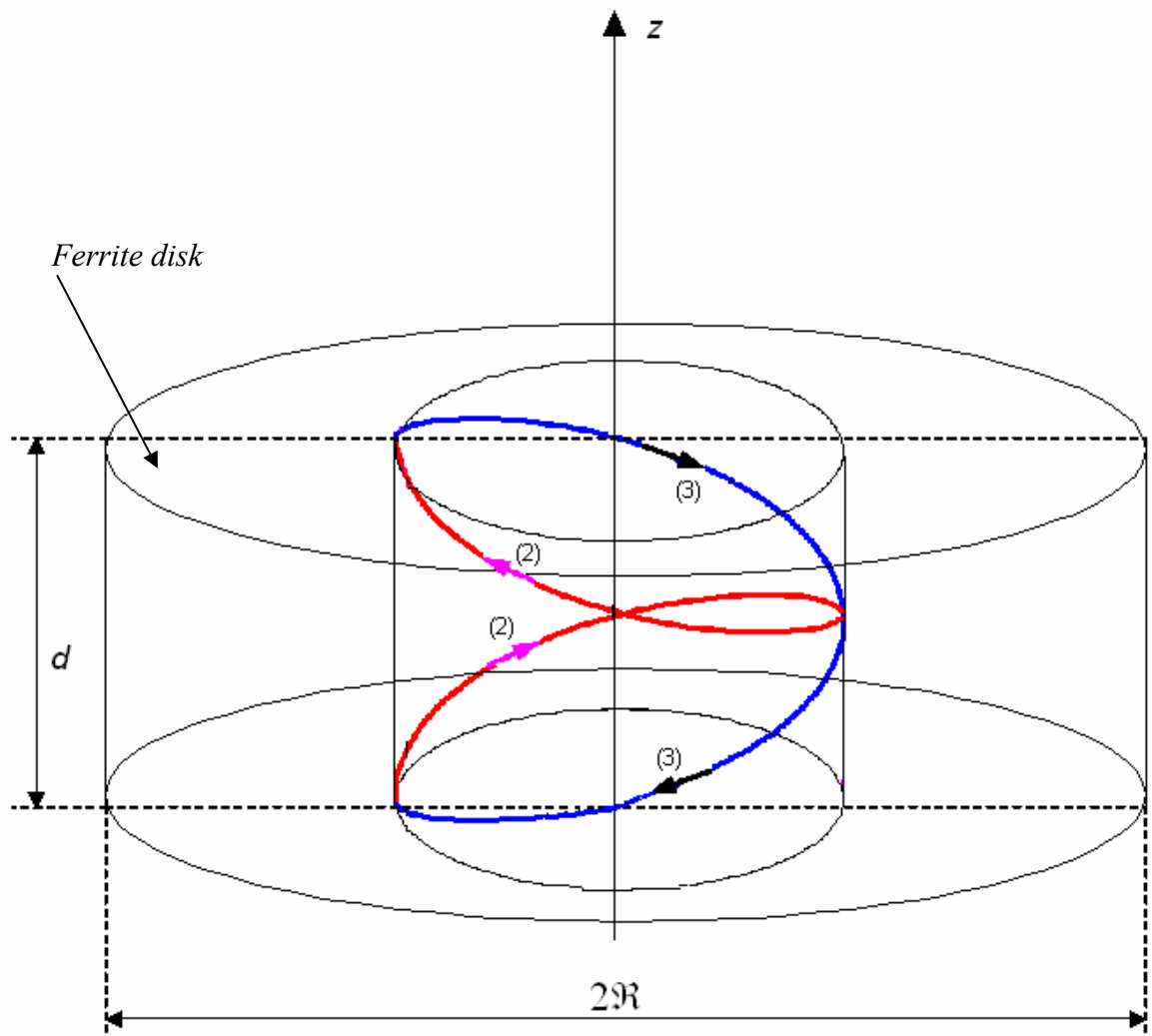

Fig 1b: The "left" resonance caused by the $\psi^{(2)} \leftrightarrow \psi^{(3)}$ interaction (arrows show directions of propagation for helical MS modes)